\documentclass[prb,aps,epsf,showpacs,twocolumn,longbibliography]{revtex4-1}
\usepackage{times}
\usepackage{graphicx}
\usepackage{float}
\usepackage{latexsym,amsmath,amssymb,bm,euscript}
\usepackage{color}
\usepackage{subfigure}
\usepackage{epstopdf}
\usepackage[colorlinks=true,linkcolor=blue,citecolor=blue,urlcolor=blue]{hyperref}
\usepackage{hyperref}

\begin{document}

\title{Global phase diagram and quantum spin liquids in spin-$1/2$ triangular antiferromagnet}
\author{Shou-Shu Gong$^1$, W. Zhu$^2$, J.-X. Zhu$^{2,3}$, D. N. Sheng$^4$, and Kun Yang$^5$}
\affiliation{$^1$National High Magnetic Field Laboratory, Florida State University, Tallahassee, FL 32310\\
$^2$Theoretical Division, T-4 and CNLS, Los Alamos National Laboratory, Los Alamos, NM 87545 \\
$^3$Center for Integrated Nanotechnologies, Los Alamos National Laboratory, Los Alamos, NM 87545\\
$^4$Department of Physics and Astronomy, California State University, Northridge, CA 91330\\
$^5$National High Magnetic Field Laboratory and Department of Physics, Florida State University, Tallahassee, FL 32306}

\begin{abstract}
We study the spin-$1/2$ Heisenberg model on the triangular lattice with the nearest-neighbor $J_1 > 0$, 
the next-nearest-neighobr $J_2 > 0$ Heisenberg interactions, and the additional scalar chiral interaction 
$J_{\chi}(\vec{S}_i \times \vec{S}_j) \cdot \vec{S}_k$ for the three spins in all the triangles using 
large-scale density matrix renormalization group calculation on cylinder geometry. With increasing 
$J_2$ ($J_2/J_1 \leq 0.3$) and $J_{\chi}$ ($J_{\chi}/J_1 \leq 1.0$) interactions, we establish a quantum 
phase diagram with the magnetically ordered $120^{\circ}$ phase, stripe phase, and non-coplanar tetrahedral 
phase. In between these magnetic order phases, we find a chiral spin liquid (CSL) phase, which is identified 
as a $\nu = 1/2$ bosonic fractional quantum Hall state with possible spontaneous rotational symmetry breaking. 
By switching on the chiral interaction, we find that the previously identified spin liquid in the $J_1 - J_2$ 
triangular model ($0.08 \lesssim J_2/J_1 \lesssim 0.15$) shows a phase transition to the CSL phase at very small 
$J_{\chi}$. We also compute spin triplet gap in both spin liquid phases, and our finite-size results suggest
large gap in the odd topological sector but small or vanishing gap in the even sector.
We discuss the implications of our results to the nature of the spin liquid phases.
\end{abstract}


\maketitle

\section{Introduction}

Quantum spin liquid (QSL) is one kind of long-range entangled states with fractionalized
quasiparticles~\cite{savary2016}. Since the proposal by P. W. Anderson, 
the concept of QSL has been playing an important role for understanding strongly correlated 
materials and unconventional superconductors~\cite{Lee2006}. Although QSLs have been pursued 
for more than two decades~\cite{Rokhsar1988,baskaran1989,Read1991,sachdev1992,yang1993,Balents2002,Yao2007},
only recently such novel states have been found in realistic spin models~\cite{Kitaev2006,
Greiter2007,thomale2009,nielsen2012,bauer2014,gong2014kagome,he2014csl,hu2015csl,meng2015,
sela2015,kumar2015,lauchli2015,hickey2016,zhu2016}, in which geometric frustration and 
competing interactions play important roles for developing spin liquid states.

One of the most promising spin liquid candidates is the antiferromagnet on the corner-sharing 
kagome lattice. Experimentally, spin liquid-like behaviors have been observed in several kagome 
materials such as herbertsmithite~\cite{mendels2007,helton2007,han2012,fu2015,norman2016}. Theoretically, 
the most extensively studied kagome model is the spin-$1/2$ kagome Heisenberg model with the 
nearest-neighbor (NN) interaction. Thanks to the recent large-scale Density Matrix Renormalization 
Group (DMRG) simulations~\cite{Yan2011, Depenbrock2012}, conventional orders have been excluded, 
leading to a QSL ground state. However, the nature of this spin liquid is still in debate. DMRG 
calculations suggest a gapped spin liquid~\cite{Yan2011, Depenbrock2012, Jiang2012nature}, seemingly 
consistent with a $Z_2$ topological order~\cite{Depenbrock2012, Jiang2012nature}. Recent tensor 
network state simulations identify the $Z_2$ topological order of the obtained variational 
wavefunction~\cite{mei2016}, but so far the four degenerate ground states of the putative $Z_2$ QSL 
have not been found in exact diagonalization (ED)~\cite{waldtmann1998,lauchli2016kagome} and DMRG calculations, 
leaving this problem open. On the other hand, variational studies based on the fermionic 
parton wavefunctions find the gapless U(1) Dirac spin liquid rather than the gapped $Z_2$ spin liquid 
with the optimized variational energy~\cite{Ran2007, Iqbal2013, Iqbal2014}. Very recently, tensor 
renormalization group~\cite{jiang2016, xiang2016} and DMRG~\cite{he2016} calculations also 
suggest the gapless spin liquid as a strong candidate. Interestingly, studies on the modified kagome 
models~\cite{messio2012,gong2014kagome,he2014csl,gong2015,bauer2014,Iqbal2015} find that the kagome 
spin liquid emerges near the phase boundaries of several ordered phases, suggesting possible strong 
competitions of the different physics in the kagome spin liquid regime. In particular, a fully gapped 
chiral spin liquid (CSL)~\cite{kalmeyer1987,wen1989csl} is found by switching on small further-neighbor
~\cite{gong2014kagome,he2014csl} or chiral interactions~\cite{bauer2014} on the NN kagome model.

Another promising spin liquid candidate is the antiferromagnet on the edge-sharing triangular lattice.
Although frustration is present in the spin-$1/2$ NN triangular model, it turns out to still exhibit
a $120^{\circ}$ antiferromagnetic order~\cite{bernu1992, sorella1999}. In the recent experiments 
on the triangular organic Mott insulators such as $\kappa$-(ET)$_2$Cu$_2$(CN)$_3$ and 
EtMe$_3$Sb[Pd(dmit)$_2$]$_2$~\cite{shimizu2003, kurosaki2005, yamashita2008, yamashita2009, yamashita2010}, 
spin liquid-like behaviors have been found. Theoretically, multi-spin exchange interactions, which can
lead to the gapless spin Bose metal with a large spinon Fermi surface~\cite{Misguich1999,lesik2005,sheng2009} and the gapless spin liquid with a quadratic band touching~\cite{mishmash2013,bieri2016}
depending on the strength of interaction,
and the space anisotropic interaction~\cite{yunoki2006, sheng2006, leon2007, white2011}
have been suggested to understand the spin-liquid behaviors in triangular materials.

Recently, a new spin liquid phase is found in the spin-$1/2$ triangular Heisenberg model with the NN 
$J_1$ and the next-nearest-neighbor (NNN) $J_2$ interactions for $0.08 \lesssim J_2/J_1 \lesssim 0.15$, 
which is sandwiched by a $120^{\circ}$ magnetic phase and a stripe magnetic order 
phase~\cite{mishmash2013,kaneko2014,campbell2015,zhuzhenyue2015,Hu2015,mcculloch2016,iqbal2016}. 
This frustrating $J_2$ interaction is considered as a possible mechanism to understand the 
spin-liquid behaviors of the newly 
synthesized triangular materials YbMgGaO$_4$~\cite{paddison2016} and Ba$_3$InIr$_2$O$_9$~\cite{dey2017}. 
For this $J_1-J_2$ model, DMRG calculations on cylinder system find the evidence of spin liquid 
including the two near-degenerate ground states in the even and odd topological sectors whose energy 
difference decays rapidly with growing cylinder width, and the fractionalized spin-$1/2$ quasiparticle 
revealed by inserting flux simulation and entanglement spectrum (ES)~\cite{zhuzhenyue2015,Hu2015,
mcculloch2016}. On the finite-size DMRG calculations, the spin triplet gap measured above the 
overall ground state (in the odd sector) is big ($\Delta_{\rm T} \sim 0.3J_1$)~\cite{zhuzhenyue2015,Hu2015}, 
seemingly consistent with a gapped spin liquid~\cite{zheng2015,lu2016}. Nonetheless, the even and odd sectors show 
some distinct features in finite-size DMRG calculation. While the odd sector shows a short correlation 
length that could be consistent with the large gap, the even sector exhibits a much larger 
one~\cite{Hu2015,mcculloch2016}, which may suggest smaller gap in the even sector.
The low-lying entanglement spectrum in the even sector shows a Dirac node like structure, 
which is suggested as an implication of gapless spinon excitations~\cite{mcculloch2016}.
The different DMRG results in the two sectors reasonably imply that {\it either} the putative 
gapped spin liquid is not yet well developed as the strong finite-size effects in numerical 
calculation, {\it or} a gapless spin liquid is possible. In the 
variational study, a U(1) Dirac gapless spin liquid indeed possesses the best variational 
energy~\cite{iqbal2016}. The nature of this spin liquid remains an open question. To shed more 
light on this spin liquid phase, the modified $J_1 - J_2$ triangular models have been 
investigated~\cite{yao2015,hu2016,lauchli2016,misumi2017}. Interestingly, the variational~\cite{hu2016} 
and ED calculations~\cite{lauchli2016} suggest a possible CSL at the neighbor of the $J_1 - J_2$ spin 
liquid, which seems to be similar to the situation in the kagome model and deserves more studies. 
Besides, the quantum phase transition between the two spin liquid phases is also far from clear.

\begin{figure}[t]
\includegraphics[width=1.0\linewidth]{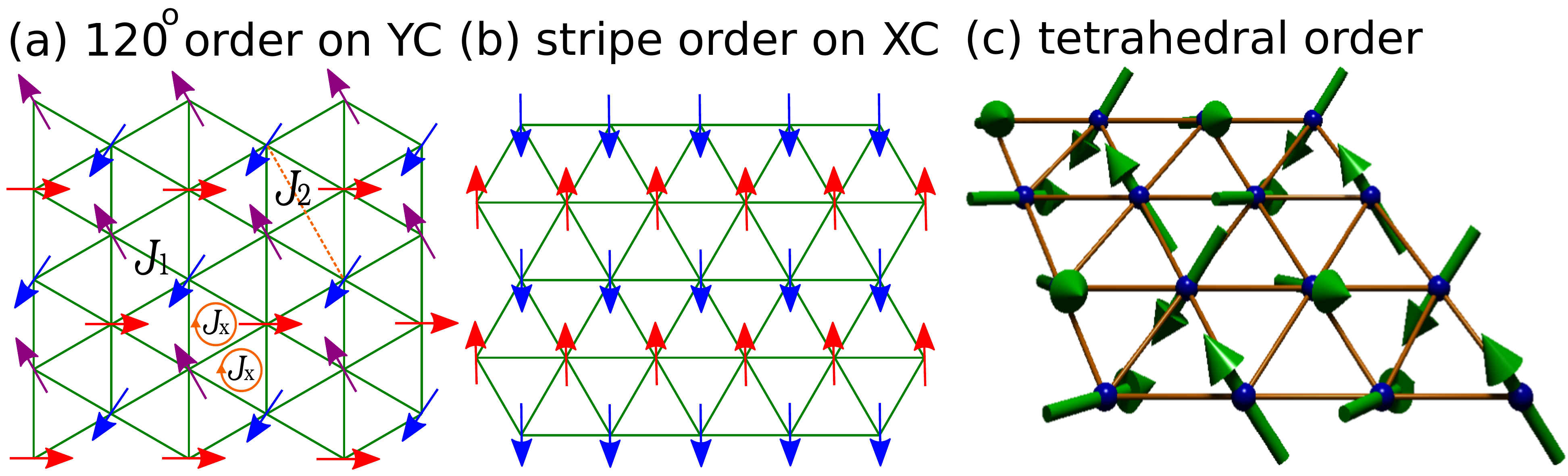}
\includegraphics[width=1.0\linewidth]{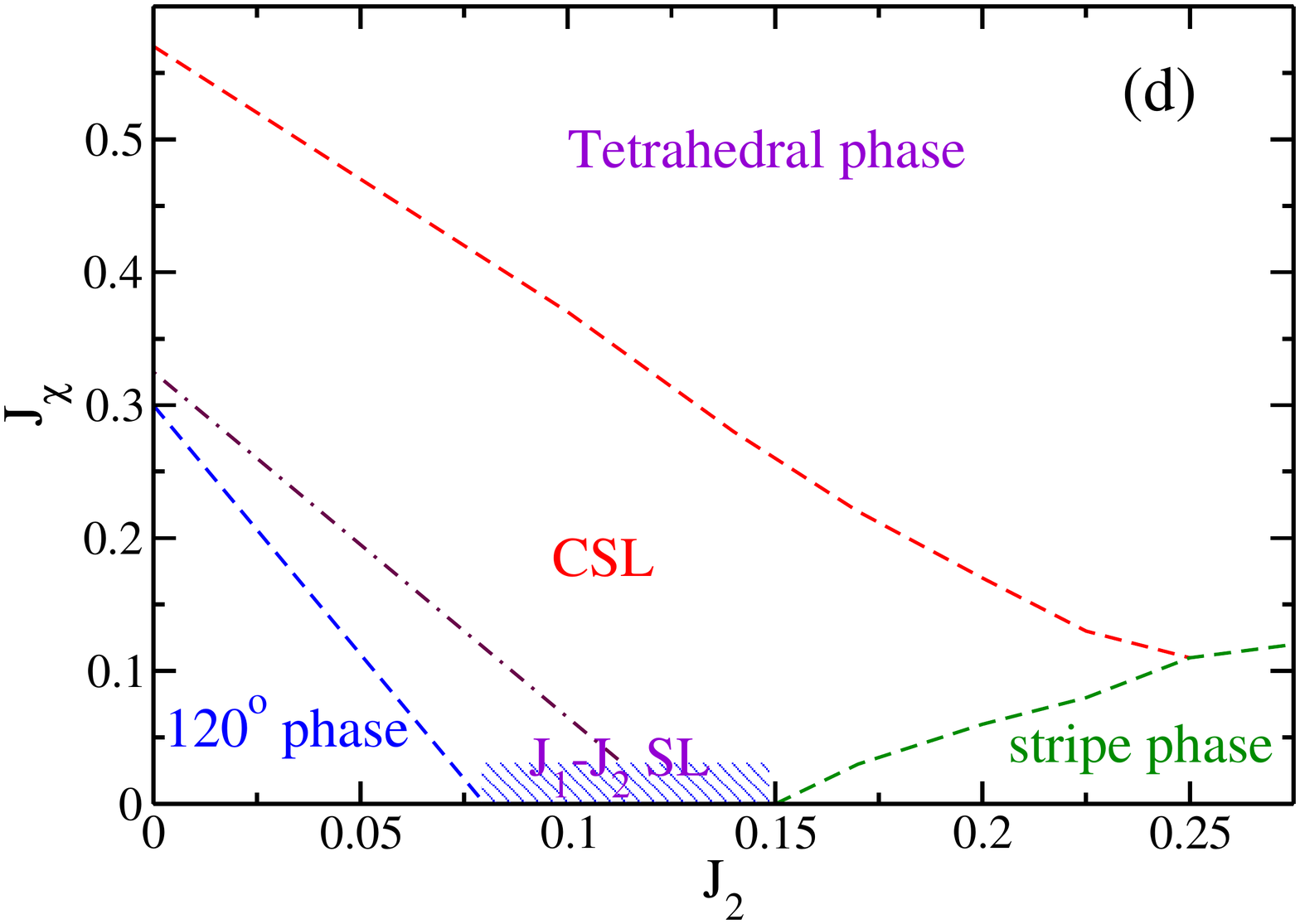}
\caption{Model Hamiltonian and quantum phase diagram of the spin-$1/2$ $J_1-J_2-J_{\chi}$ Heisenberg model
on the triangular lattice. (a) and (b) are the schematic figures of the $120^{\circ}$ and the stripe
magnetic order on the XC and YC cylinders. The triangular model has the nearest-neighbor $J_1$, 
next-nearest-neighbor $J_2$, and three-spin scalar chiral interaction $J_{\chi}$. For all the triangles, 
the chiral interactions have the same chirality direction. (c) Tetrahedral magnetic order on the triangular 
lattice. This order has four sublattices with spins pointing toward the corners of a tetrahedron. 
(d) Quantum phase diagram of the model with growing $J_2$ and $J_{\chi}$. The model shows the $120^{\circ}$ 
magnetic order phase, $J_1 - J_2$ spin liquid ($J_1 - J_2$ SL) phase, stripe magnetic order phase, 
chiral spin liquid (CSL) phase, and tetrahedral phase. The phase boundaries (dashed lines) are obtained 
by measuring magnetic order parameter and spin correlation function. The dotdashed line is the classical
phase boundary between the $120^{\circ}$ magnetic order and the tetrahedral order.
}
\label{fig:phase}
\end{figure}

In this article, we study the spin-$1/2$ $J_1-J_2$ Heisenberg model on the triangular lattice with 
additional time-reversal symmetry (TRS) breaking chiral interaction $J_{\chi}$ using DMRG simulations. 
The model Hamiltonian is given as
\begin{equation}
 H = J_1 \sum_{\langle i,j \rangle} \vec{S}_i \cdot \vec{S}_j 
   + J_2 \sum_{\langle\langle i,j \rangle\rangle} \vec{S}_i \cdot \vec{S}_j
   + J_{\chi} \sum_{\bigtriangleup /\bigtriangledown} (\vec{S}_i \times \vec{S}_j)\cdot \vec{S}_k, \nonumber
\label{eq:ham}
\end{equation}
where $J_1$ and $J_2$ denote the NN and the NNN interactions, respectively. The scalar chiral 
interaction $J_{\chi}$ has the same magnitude for all the up ($\bigtriangleup$) and down 
($\bigtriangledown$) triangles, and the three sites $i,j,k$ for $J_{\chi}$ follow the clockwise 
order in all the triangles as shown in Fig.~\ref{fig:phase}(a). Physically, the scalar chiral interaction 
$J_{\chi}$ term can be induced in the Hubbard model with large $U$ in a magnetic 
field~\cite{diptiman1995,lesik2006}. Starting from the Hubbard model, a $t/U$ ($t$ and $U$ 
are the hopping and interaction respectively) expansion to the second order at half filling 
gives the effective chiral interaction $J_{\chi} (\vec{S}_i \times \vec{S}_j)\cdot \vec{S}_k$ 
with $J_{\chi} \sim \Phi t^3/U^2$, where $\Phi$ is the magnetic flux enclosed by the triangle.
We take $J_1 = 1.0$ as the energy scale. Using DMRG simulation, we obtain a quantum phase 
diagram as shown in Fig.~\ref{fig:phase}(d). Besides the $120^{\circ}$ N\'eel phase, 
the stripe phase, and the time-reversal invariant spin liquid in the $J_1 - J_2$ model (here 
we denote it as $J_1 - J_2$ SL), we find a large regime of the non-coplanar tetrahedral 
order for large $J_{\chi}$, whose spin configuration is shown in Fig.~\ref{fig:phase}(c). 
Below the tetrahedral phase for $J_2 \lesssim 0.25$, we identify a CSL as the $\nu = 1/2$ bosonic
fractional quantum Hall state by observing the gapless chiral edge mode. The strong nematic order of bond 
energy suggests a possible spontaneous lattice rotational symmetry breaking and implies an emergent 
nematic CSL. By studying the spin triplet gap and entanglement spectrum, we observe a transition 
from the $J_1 - J_2$ SL to the CSL at small chiral interaction. While we find a large spin
triplet gap above the overall ground state (in the odd sector) in the CSL phase, the small triplet gap in 
the even sector suggests that on our studied system size the topological nature in the even sector 
may not have been fully developed. A possible reason is that this CSL regime generated by increasing 
$J_{\chi}$ is near the phase boundaries from the CSL to the neighboring phases. In the $J_1 - J_2$ triangular 
model, the triplet gap in the even sector seems to be vanished, which could be consistent with the 
larger correlation length found in DMRG calculation~\cite{Hu2015, mcculloch2016} and may suggest a 
possible gapless spin liquid~\cite{iqbal2016}, which deserves more studies.

We study the system with cylinder geometry using DMRG~\cite{white1992} with spin rotational SU(2) 
symmetry~\cite{mcculloch2002}. We choose two geometries that have one lattice direction 
parallel to either the $x$ axis (XC) or the $y$ axis (YC), as shown in Figs.~\ref{fig:phase}(a)-(b). 
These  cylinders are denoted as XC(YC)$L_y$-$L_x$, where $L_y$ and $L_x$ are the numbers of site 
along the two directions. To study the phase diagram and characterize the CSL phase, we perform
calculations on the systems with $L_y$ up to $8$ and $10$. We keep up to $4000$ SU(2) states to 
obtain accurate results with the truncation error less than $10^{-5}$ in most calculations.

\begin{figure}[t]
\includegraphics[width=.7\linewidth]{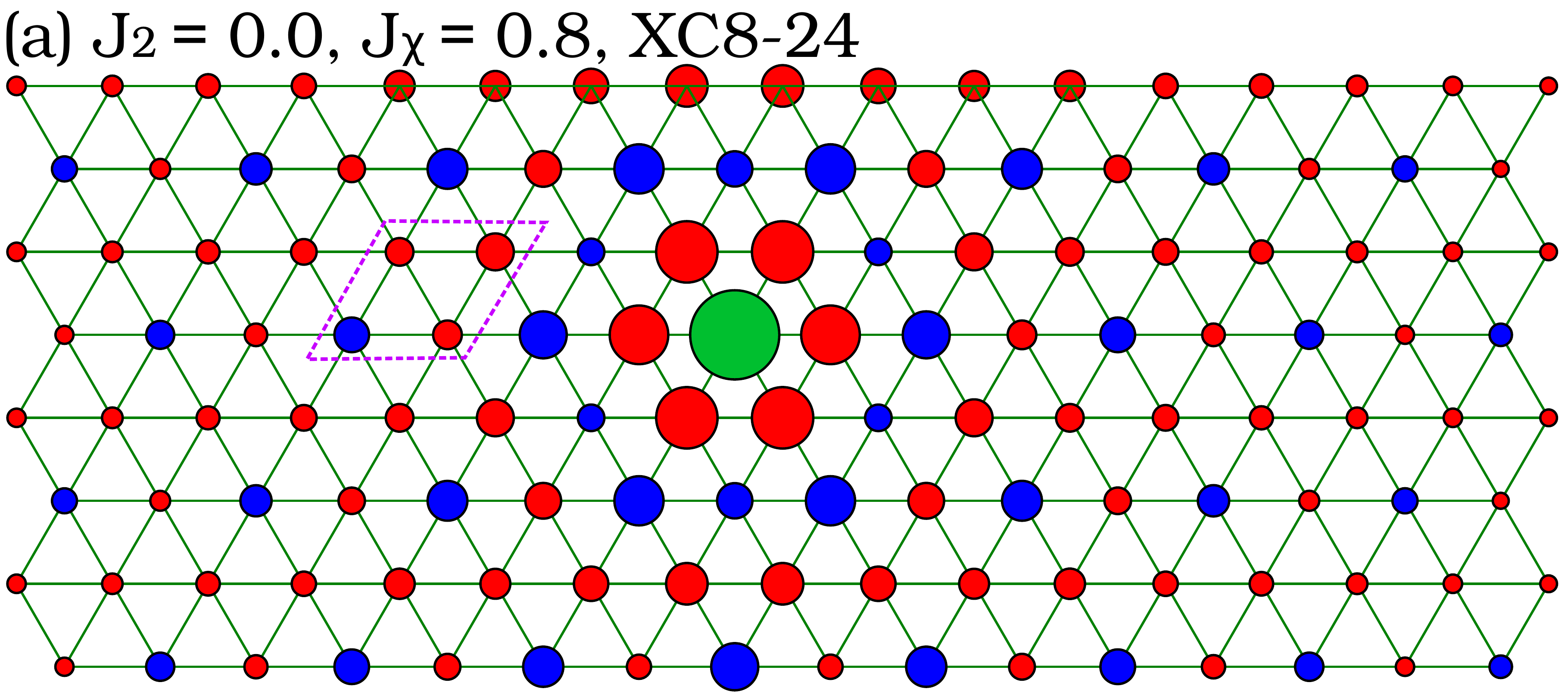}
\includegraphics[width=.7\linewidth]{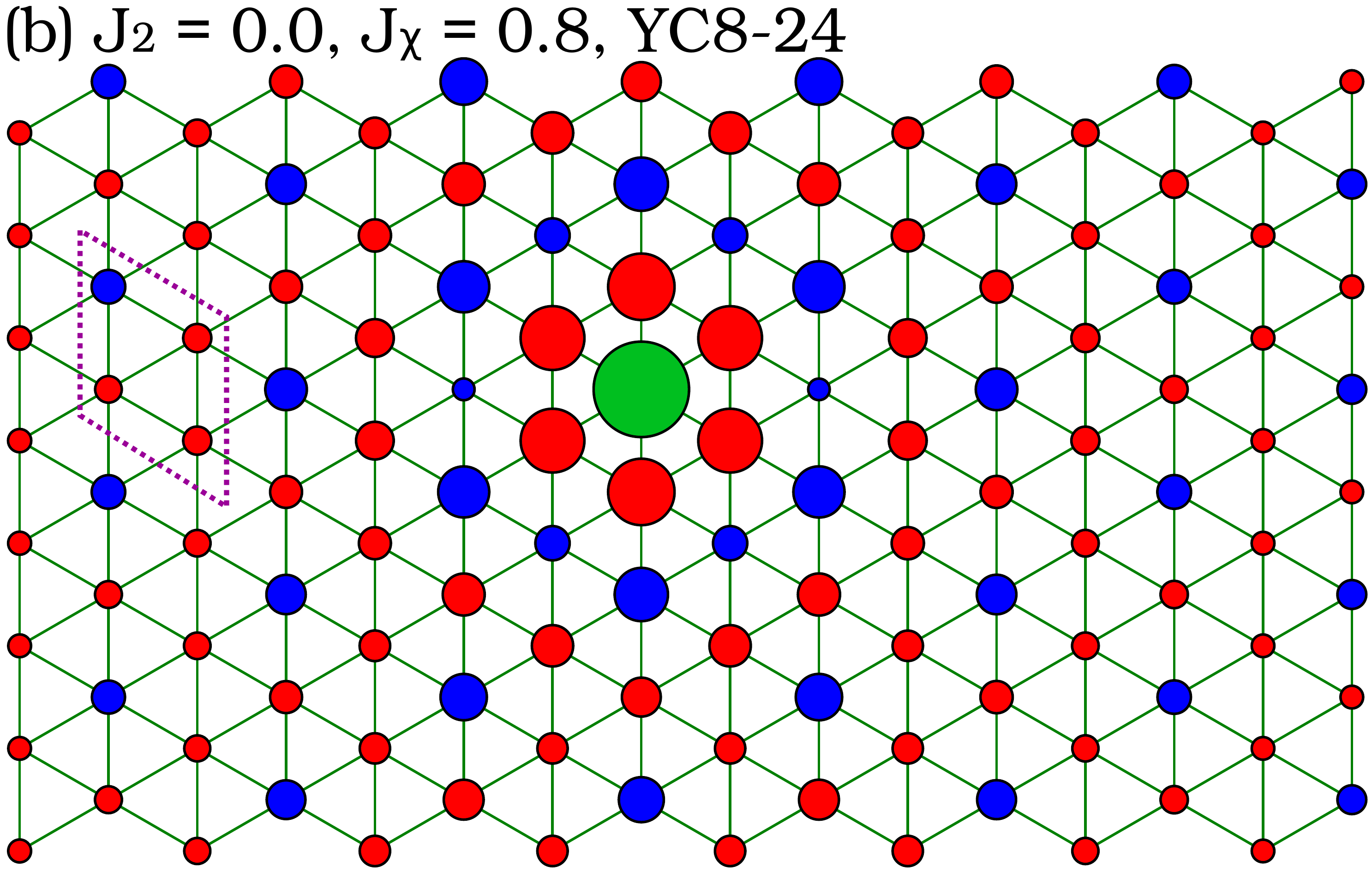}
\includegraphics[width=0.49\linewidth]{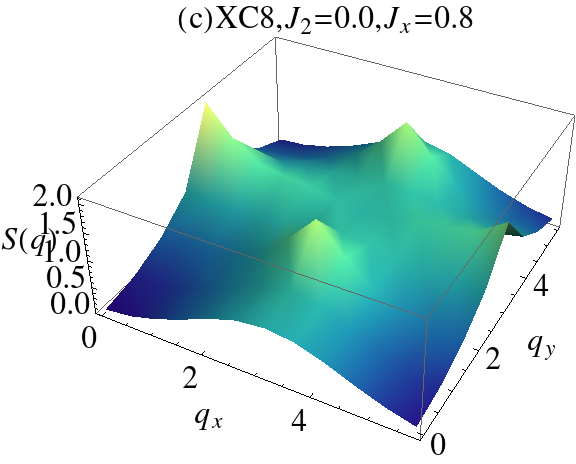}
\includegraphics[width=0.49\linewidth]{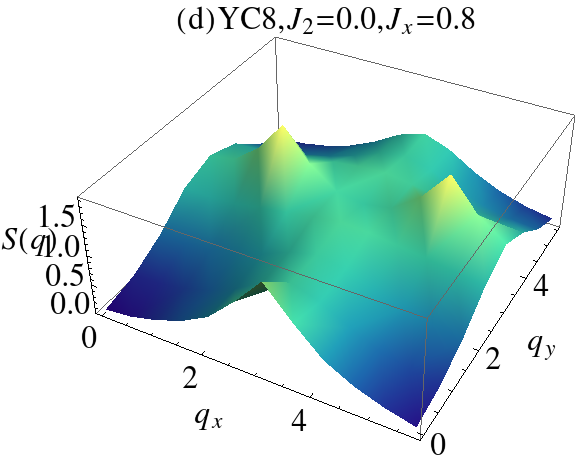}
\caption{Tetrahedral magnetic order for $J_2 = 0.0, J_{\chi} = 0.8$.
(a) and (b) are the spin correlation functions $\langle \vec{S}_i \cdot \vec{S}_j \rangle$ 
in real space for the XC8-24 and YC8-24 cylinders, respectively. The green circle denotes
the reference spin. The blue and red circles denote the positive and negative correlations, 
respectively. The magnitude of correlations are proportional to the square of circle area. 
The spin correlations on both geometries are consistent with the tetrahedral magnetic 
order with the unit cell shown by the dashed diamond.
(c) and (d) are the corresponding spin structure factor $S(\vec{q})$ of (a) and (b).
On the XC and YC cylinders, the tetrahedral state has the structure factor peaks at 
$\vec{q} = (0,\pi), (\pi,\pi/2)$ and $(\pi/2,\pi)$, respectively.}
\label{fig:tetra_spin}
\end{figure}

\section{Tetrahedral order and 120$^{\circ}$ order}
\label{sec:order}

For $J_2 = 0.0$, the triangular model has a coplanar $120^{\circ}$ magnetic order at
$J_{\chi} = 0.0$~\cite{bernu1992, sorella1999, zheng2006, white2007}. In the large 
$J_{\chi}$ limit, a classical spin analysis finds a tetrahedral magnetic state with the 
spins of the four sublattices pointing toward the corners of a tetrahedron~\cite{messio2011} 
(see Fig.~\ref{fig:phase}(b)). In the classical picture, the $120^{\circ}$ ordered state has 
the energy per site $E_{120} = - 3J_1/2 + 3J_2$, the stripe ordered state has the energy 
$E_{\rm stripe} = -J_1 - J_2$, and the tetrahedral ordered state has the energy 
$E_{\rm tetra} = -J_1 - J_2 - 8\sqrt{3}J_{\chi}/9$. Thus we can get a classical phase
diagram in the $J_2 - J_{\chi}$ plane. In Fig.~\ref{fig:phase}(d), the dotdashed line denotes
the classical phase boundary between the $120^{\circ}$ and the tetrahedral phase. 
For $J_{\chi} = 0.0, J_2 > 0.125$, the stripe state and the tetrahedral state have the
degenerate energy. By switching on the chiral interaction, the tetrahedral state immediately
gets the lower energy. In quantum model, we first investigate whether this tetrahedral 
order could survive for the spin-$1/2$ system with strong quantum 
fluctuations. In Figs.~\ref{fig:tetra_spin}(a)-(b), we demonstrate the spin correlations 
$\langle \vec{S}_i \cdot \vec{S}_j \rangle$ for $J_2 = 0.0, J_{\chi} = 0.8$ on both the 
XC8-24 and YC8-24 cylinders. The dashed diamonds denote the unit cell of the spin correlation, 
which is consistent with the four sublattice structure of the tetrahedral order. The spin 
correlation functions decay quite slowly in both systems, indicating an established 
magnetic order. In Figs.~\ref{fig:tetra_spin}(c)-(d), we show the corresponding spin structure factor 
$S(\vec{q}) = \frac{1}{N}\sum_{i,j}\langle \vec{S}_i \cdot \vec{S}_j \rangle e^{i\vec{q}\cdot (\vec{r}_i 
- \vec{r}_j)}$ of the tetrahedral state on both cylinder geometries, which has the ordering 
peaks at $\vec{q} = (0,\pi), (\pi,\pi/2)$ and $(\pi/2,\pi)$ on the XC8 and YC8 cylinders, respectively.

\begin{figure}[t]
\includegraphics[width=0.49\linewidth]{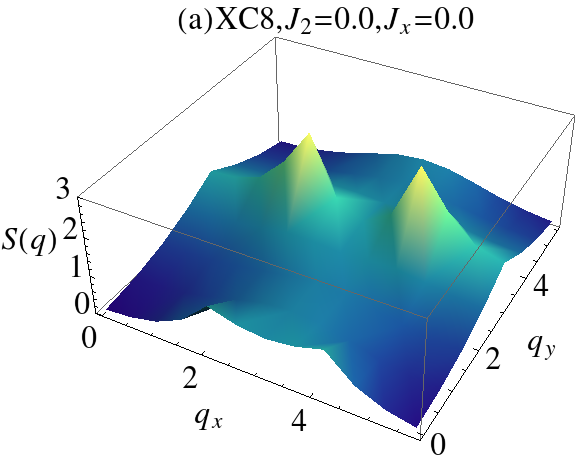}
\includegraphics[width=0.49\linewidth]{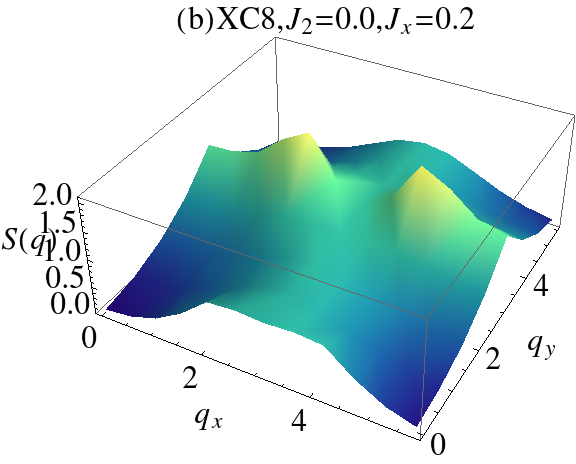}
\includegraphics[width=0.49\linewidth]{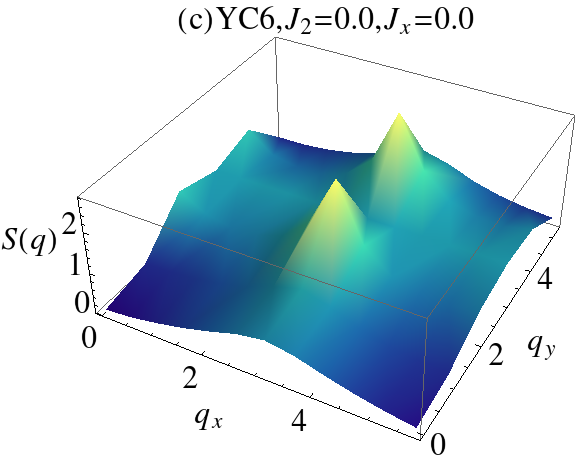}
\includegraphics[width=0.49\linewidth]{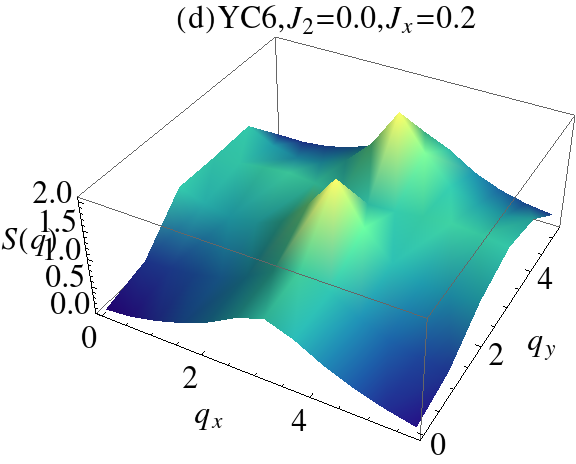}
\caption{Spin structure factor $S(\vec{q})$ in the $120^{\circ}$ magnetic order phase.
(a) and (b) are obtained from the middle XC8-12 sites on the XC8-30 cylinder, and
(c) and (d) are obtained from the middle YC6-12 sites on the YC6-30 cylinder.
Both XC and YC systems exhibit the characteristic peak of the $120^{\circ}$ order at 
$\vec{q} = (2\pi/3, \pi)$ and $(\pi, 2\pi/3)$, respectively.
}
\label{fig:sq}
\end{figure}

For small $J_{\chi}$ interaction we expect the $120^{\circ}$ magnetic order.
In Fig.~\ref{fig:sq}, we show the spin structure factor for $J_{\chi} = 0.0, 0.2$ 
on both the XC8 and YC6 cylinders. For a finite $J_{\chi} = 0.2$,
the characteristic peak of the $120^{\circ}$ order is still very stable, 
indicating the dominant three-sublattice spin structure.

\section{Quantum spin liquids}

\subsection{Chiral spin liquid}

For showing our results of the chiral spin liquid phase, we choose the parameters 
with fixed $J_{2} = 0.1$, where the system is in the $J_1 - J_2$ SL in the absence 
of the chiral interaction~\cite{campbell2015,zhuzhenyue2015,Hu2015,mcculloch2016,iqbal2016}
(we have also studied other $J_2$ such as $J_2 = 0.125$, which gives the same results).
In our DMRG simulation of spin liquid phase on cylinder geometry, we control the even/odd 
parity of spinon flux in the ground state by removing or adding a spin-$1/2$ on each open edge 
of cylinder~\cite{Yan2011, zhuzhenyue2015, Hu2015}.

\begin{figure}[t]
\includegraphics[width=1.0\linewidth]{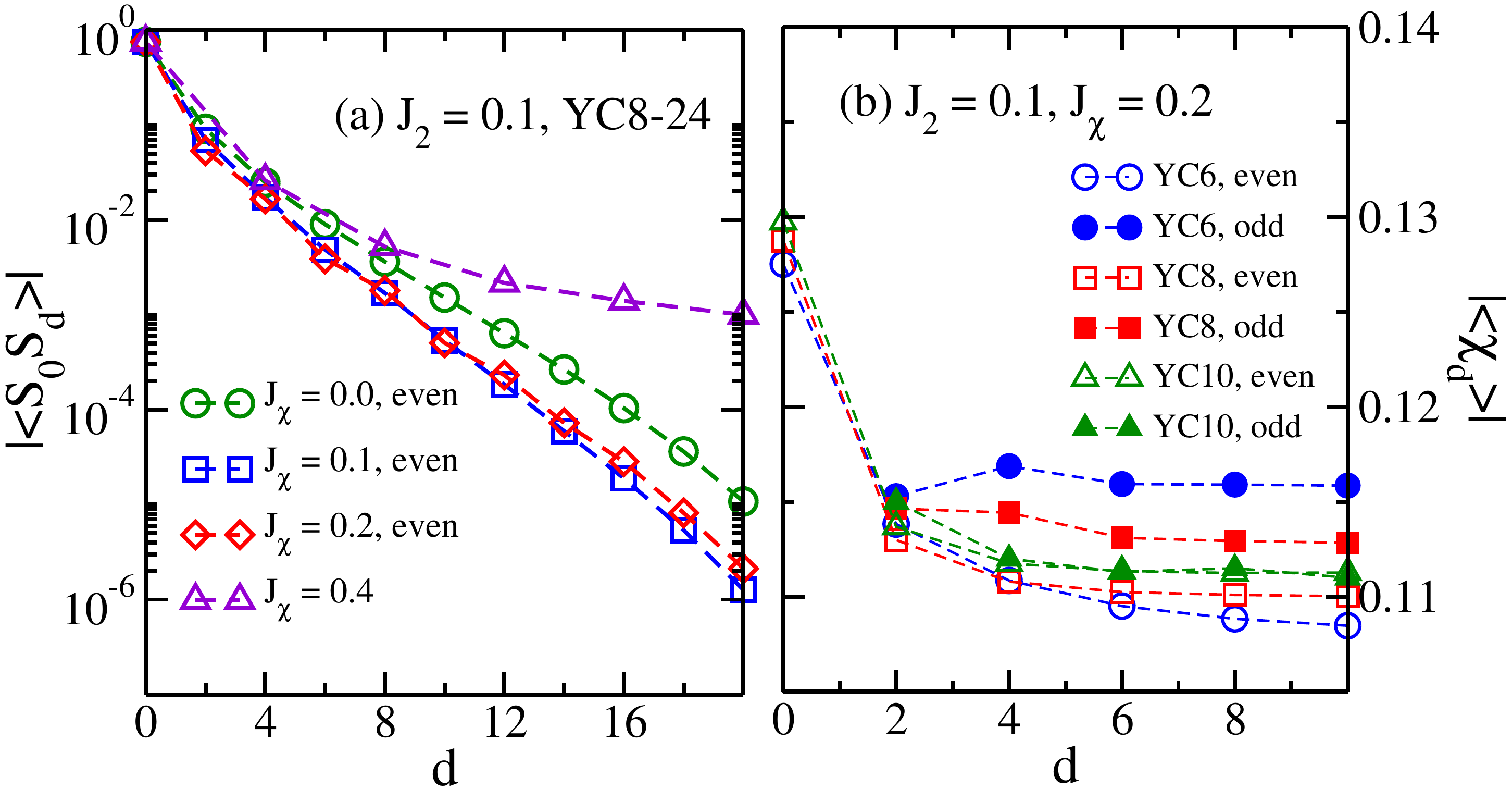}
\caption{Vanished magnetic order and non-zero chiral order in the CSL phase.
(a) Log-linear plot of spin correlations for $J_2 = 0.1$ and different $J_{\chi}$ on the YC8-24 cylinder.
(b) Distance dependence of the chiral order of triangle $|\langle \chi_{d} \rangle|$ from the open 
boundary to the bulk of cylinder for $J_2 = 0.1, J_{\chi} = 0.2$  for  different systems.}
\label{csl}
\end{figure}

\begin{figure}[t]
\includegraphics[width=.7\linewidth]{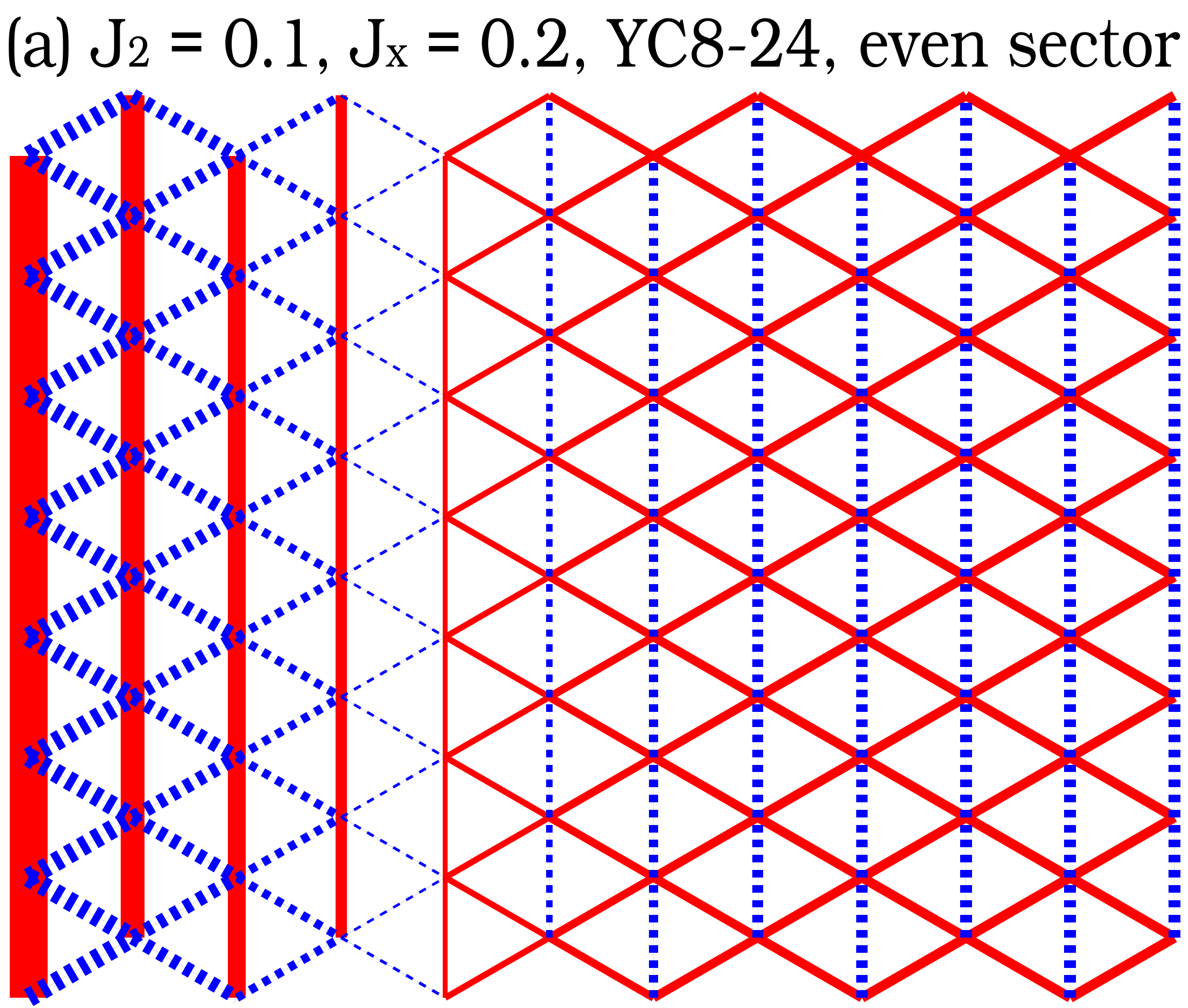}
\includegraphics[width=.7\linewidth]{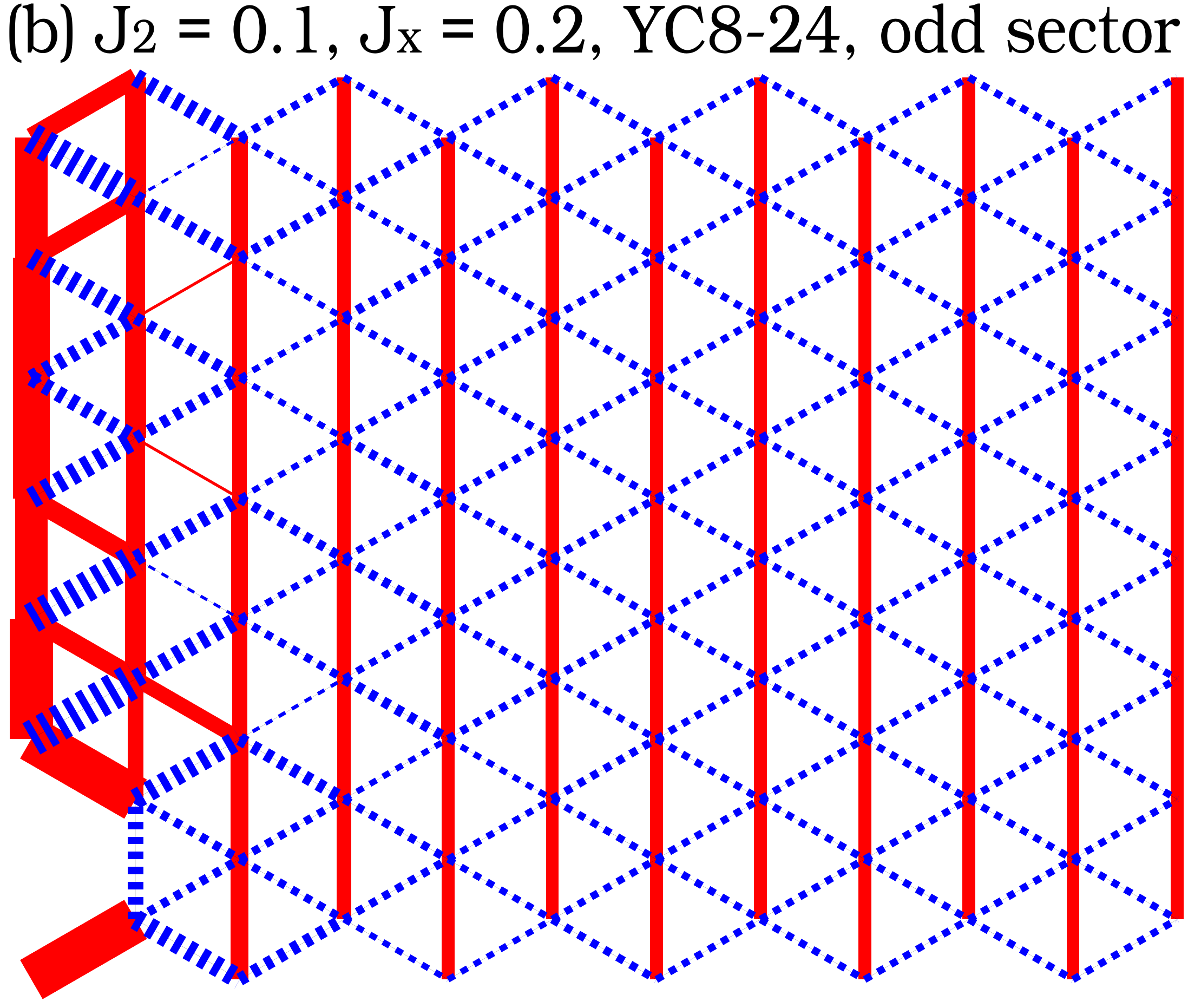}
\includegraphics[width=.7\linewidth]{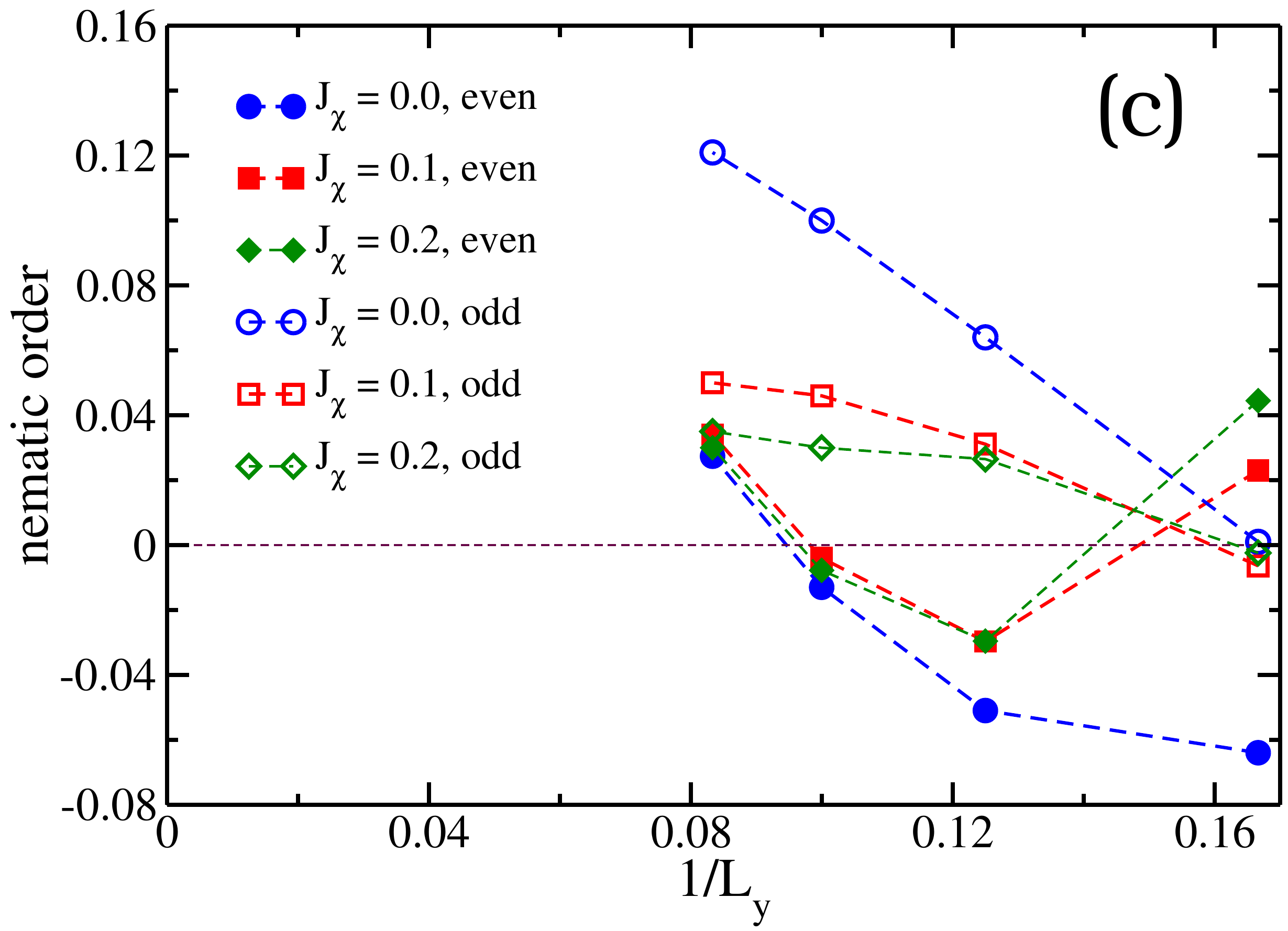}
\caption{Strong bond nematicity in the CSL phase. (a) and (b) are the nearest-neighbor bond energy 
$\langle \vec{S}_i \cdot \vec{S}_j \rangle$ for $J_2 = 0.1, J_{\chi} = 0.2$ on the YC8-24 cylinder in the 
even and odd sectors. The left half systems are shown here. The odd sector in (b) is obtained by removing
one site in each boundary of cylinder. In both figures, all the bond energy have subtracted a constant. 
The red solid and blue dashed bonds denote the negative and positive bond energies after subtraction. 
(c) Cylinder width dependence of bond nematic order for $J_2 = 0.1$ in both the even and odd sectors 
on the YC ($L_y = 6, 8, 10, 12$) cylinder. The nematic order is defined as the difference between the zigzag 
and the vertical bond energy. The data for $J_{\chi} = 0.0$ at $L_y = 6,8,10$ have been shown in 
Ref.~\onlinecite{Hu2015}.}
\label{nematic}
\end{figure}

We first exclude the conventional orders in the CSL phase. We show the spin correlations in 
Fig.~\ref{csl}(a), where the correlations in the CSL phase decay faster than those in the 
$J_1 - J_2$ SL, indicating the vanished magnetic order. In Fig.~\ref{csl}(b), we plot the 
triangle chiral order $\langle \chi_{\Delta_i} \rangle$ along the $x$ direction of cylinder. 
Different from the decayed chiral order in the $J_1 - J_2$ triangular model~\cite{Hu2015,mcculloch2016}, 
here it rapidly converges to finite value and seems to be robust with increasing system width. 
For $L_y = 6, 8, 10$, the chiral orders in both sectors approach to each other, which agrees 
with the consistent local orders in different sectors of gapped spin liquid. Similar to the 
$J_1 - J_2$ SL~\cite{zhuzhenyue2015, Hu2015, mcculloch2016}, lattice translational symmetry is also 
preserved in the CSL phase, which we do not discuss in detail but show an example in 
Fig.~\ref{nematic}.

\begin{figure}[t]
\includegraphics[width=1.0\linewidth]{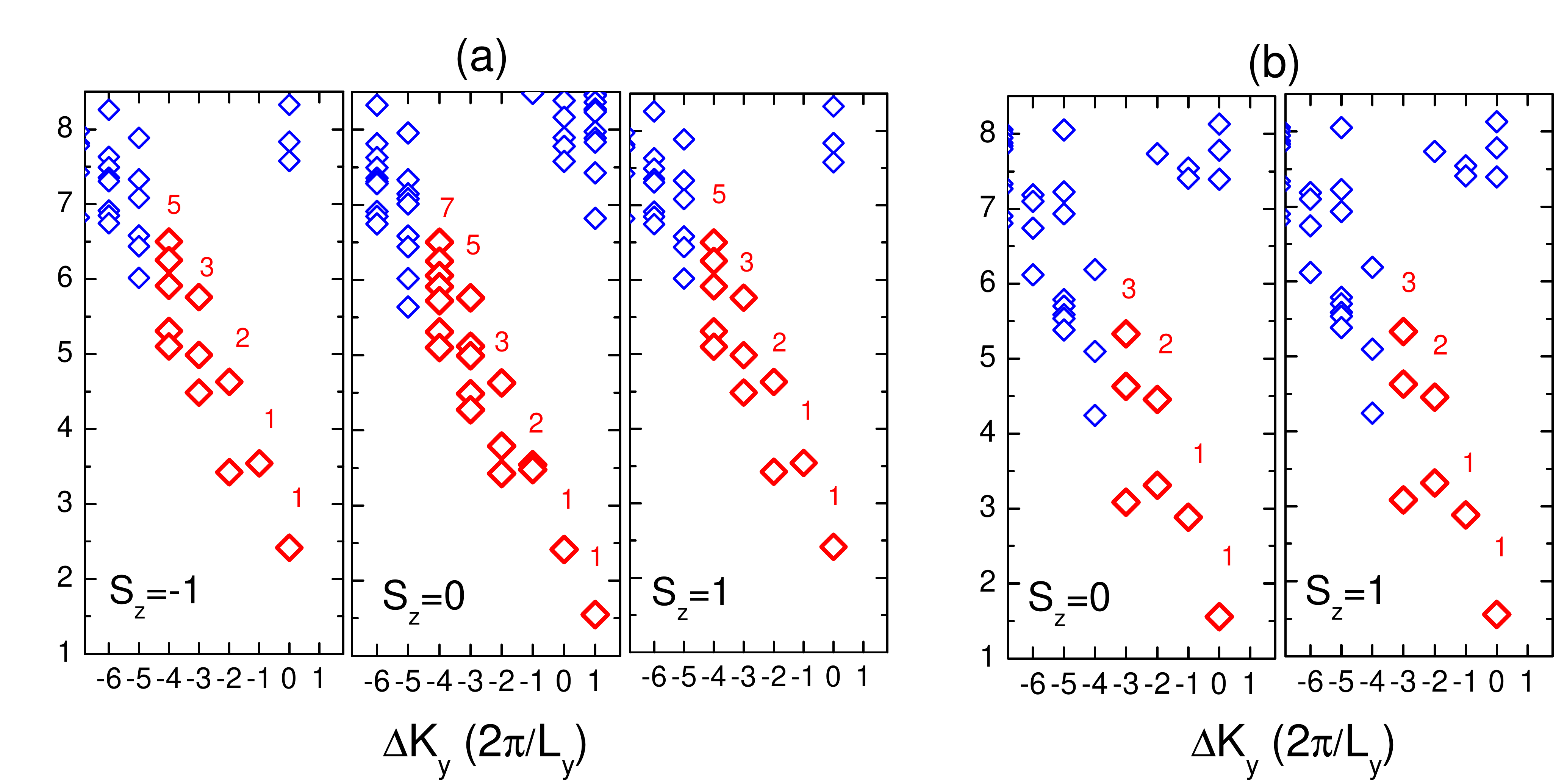}
\caption{Characterizing the CSL phase through the entanglement spectrum.
Entanglement spectra of the ground states in the even (a) and odd (b) sectors for 
$J_2 = 0.1$, $J_{\chi} = 0.2$ on the $L_y = 8$ cylinder. $\lambda_i$ is the eigenvalue
of the reduced density matrix obtained by bipartiting the cylinder system. The numbers denote
the near degenerate pattern $\{1, 1, 2, 3, 5, 7,\cdots\}$ of the low-lying spectrum with different
relative momentum quantum number $\Delta k_y$ in each spin-$S^z$ sector. }
\label{csl_spec}
\end{figure}

\begin{figure}[t]
\includegraphics[width=1.0\linewidth]{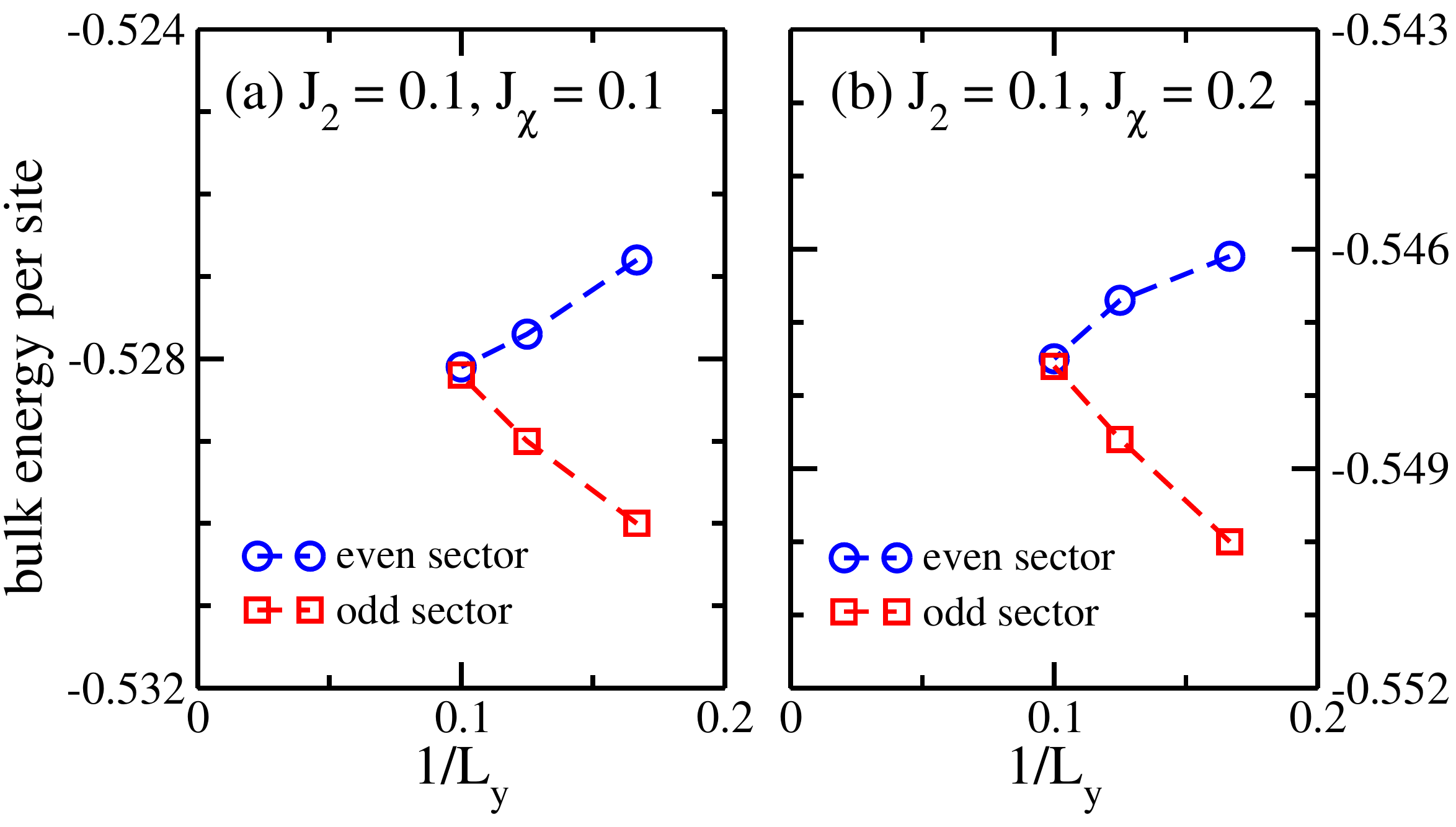}
\caption{Near-degenerate ground states in the CSL phase. 
Size dependence of the ground-state energy for $J_2 = 0.1, J_{\chi} = 0.1, 0.2$
in both the even and odd sectors on the YC6, YC8, and YC10 cylinders.
}\label{degeneracy}
\end{figure}

Next, we characterize the CSL by identifying the conformal field theory (CFT) that describes the gapless
edge excitations through entanglement spectrum~\cite{li2008}. Since this strategy was proposed~\cite{li2008}, 
the ES has been shown as a powerful tool to identify different topological orders with edge 
states~\cite{cincio2013,zaletel2013}. The nature of the CSLs in the kagome and honeycomb spin models have 
been characterized using ES~\cite{bauer2014, he2014csl, hickey2016, zhu2016}.
In Fig.~\ref{csl_spec}, we show the ES of the reduced density matrix for half the cylinder in both sectors. 
By tracing out half of the degrees of freedom for the density matrix by bipartiting the cylinder, 
we obtain the reduced density matrix and its eigenvalues $\lambda_i$. We focus on the leading eigenvalues by
showing $-\ln \lambda_i$ in Fig.~\ref{csl_spec}. The spectra are labeled by the quantum number total spin 
$S^z$ and relative momentum quantum number along the $y$ direction $\Delta k_y$~\cite{cincio2013, zaletel2013}. 
The leading spectra in both sectors have the degeneracy pattern $\{1, 1, 2, 3, 5, 7,\cdots\}$ with
increasing $\Delta k_y$ in each $S^z$ sector, which follow the chiral $SU(2)_1$ Wess-Zumino-Witten CFT 
theory of the $\nu = 1/2$ fractional quantum Hall state~\cite{cft}. The spectra of the even and odd sectors
are symmetric about $S^z=0$ and $1/2$ respectively, indicating a spin-$1/2$ at each end of cylinder in the 
odd (spinon) sector. The similar degeneracy pattern of entanglement spectra have also been found in the CSLs
in the kagome and honeycomb spin models~\cite{bauer2014, he2014csl, hickey2016, zhu2016}.
We further demonstrate the near-degenerate ground states in the CSL phase. 
In Fig.~\ref{degeneracy}, we show the bulk energies in both the even and odd sectors, where the energy difference
drops fast with increasing $L_y$. For $L_y = 10$, the energy difference is vanishing small, 
in consistent with the near-degenerate ground states in the two sectors. 

For the $J_1 - J_2$ SL, DMRG calculations find the large lattice nematic order in the odd sector
(defined as the energy difference between the zigzag and vertical 
bonds)~\cite{zhuzhenyue2015,Hu2015,mcculloch2016}, suggesting a spin liquid with possible spontaneous 
rotational symmetry breaking. However, the nematic order in the even sector exhibits the opposite 
nematic pattern from the odd sector, which seems to approach vanishing with increasing 
$L_y = 6, 8,10$~\cite{zhuzhenyue2015, Hu2015, mcculloch2016}. In the CSL phase, we also calculate 
the NN bond energy $\langle \vec{S}_i \cdot \vec{S}_j \rangle$ on the YC cylinder 
as shown in Figs.~\ref{nematic}(a)-(b). On the YC8 cylinder, we also find the strong bond 
energy anisotropy and the nematic patterns are different in the two sectors. For studying the nematic 
order, we show the nematic order on different cylinders in Fig.~\ref{nematic}(c) with a comparison 
to the data for $J_{\chi} = 0.0$~\cite{zhuzhenyue2015, Hu2015, mcculloch2016}. For $L_y = 6,8,10$, 
while the nematic order in the odd sector keeps growing with $L_y$; in the even sector it also 
appears to approach zero. The overall behaviors of the nematic order in the CSL are consistent with 
those in the $J_1 - J_2$ SL. We notice that the nematicity for the even sector in Fig.~\ref{nematic}(c)
shows a tendency to become positive with growing $L_y$. To shed more light on the nature of the 
nematicity, we calculate the bond energy for $L_y = 12$ by keeping the SU(2) states up to $6000$. 
We find that while the nematic order in the odd sector shows a consistent behavior, the order
in the even sector changes the pattern to that of the odd sector on the YC12 cylinder. 
Our results imply that in both the $J_1 - J_2$ SL and the CSL, the even sector may also host a nematic order 
in large size, suggesting possible nematic spin liquids. While CSL has been discovered in several 
spin models, the nematic CSL with coexisting topological order and nematic order has not been reported 
in a microscopic spin model as far as we know. In a strong-coupling perspective, a nematic FQH may 
be viewed as a partially melted solid, where the nematic FQH is proximate to the phase with broken 
translational and rotational symmetries~\cite{fradkin2010}. If the translational order is melted 
by tuning external parameter but nematic order is preserved, a nematic FQH might be obtained. 
Here, the CSL in the triangular model emerges at the neighbor of a stripe phase, which breaks 
translational and rotational symmetries. The possible nematicity of the CSL might be understood 
as a partially melted stripe order.

\subsection{Transition between the two spin liquids}

\begin{figure}[t]
\includegraphics[width=0.9\linewidth]{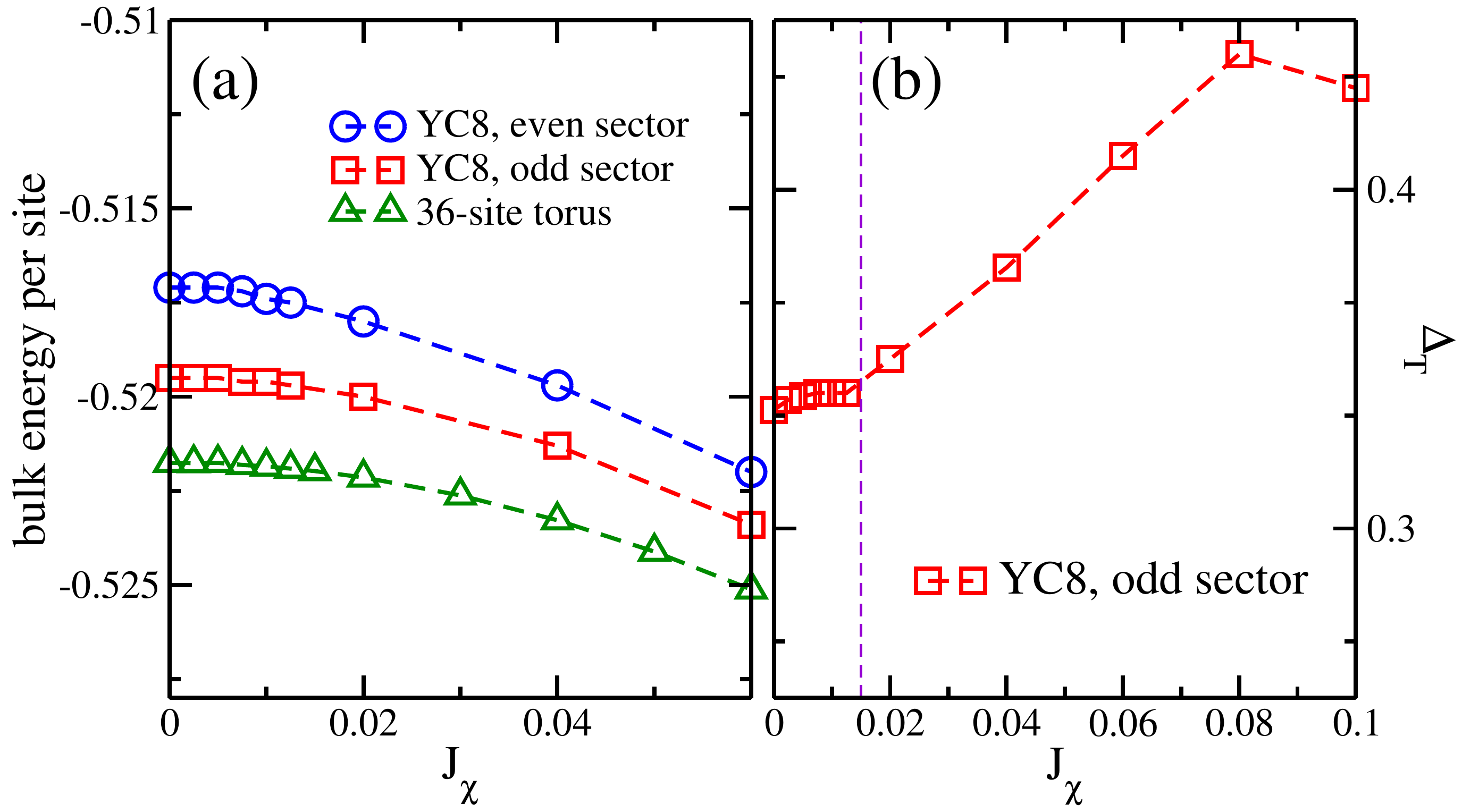}
\caption{Ground-state energy (a) and spin triplet gap $\Delta_{\rm T}$ (b) versus $J_{\chi}$ for $J_{2} = 0.1$.
The spin triplet gap is obtained by sweeping the middle $8 \times 16$ sites (total system size is YC8-24) in 
the total spin $S = 1$ sector based on the ground state with the lowest energy in the odd sector.
}\label{transition}
\end{figure}

\begin{figure}[t]
\includegraphics[width=.45\linewidth]{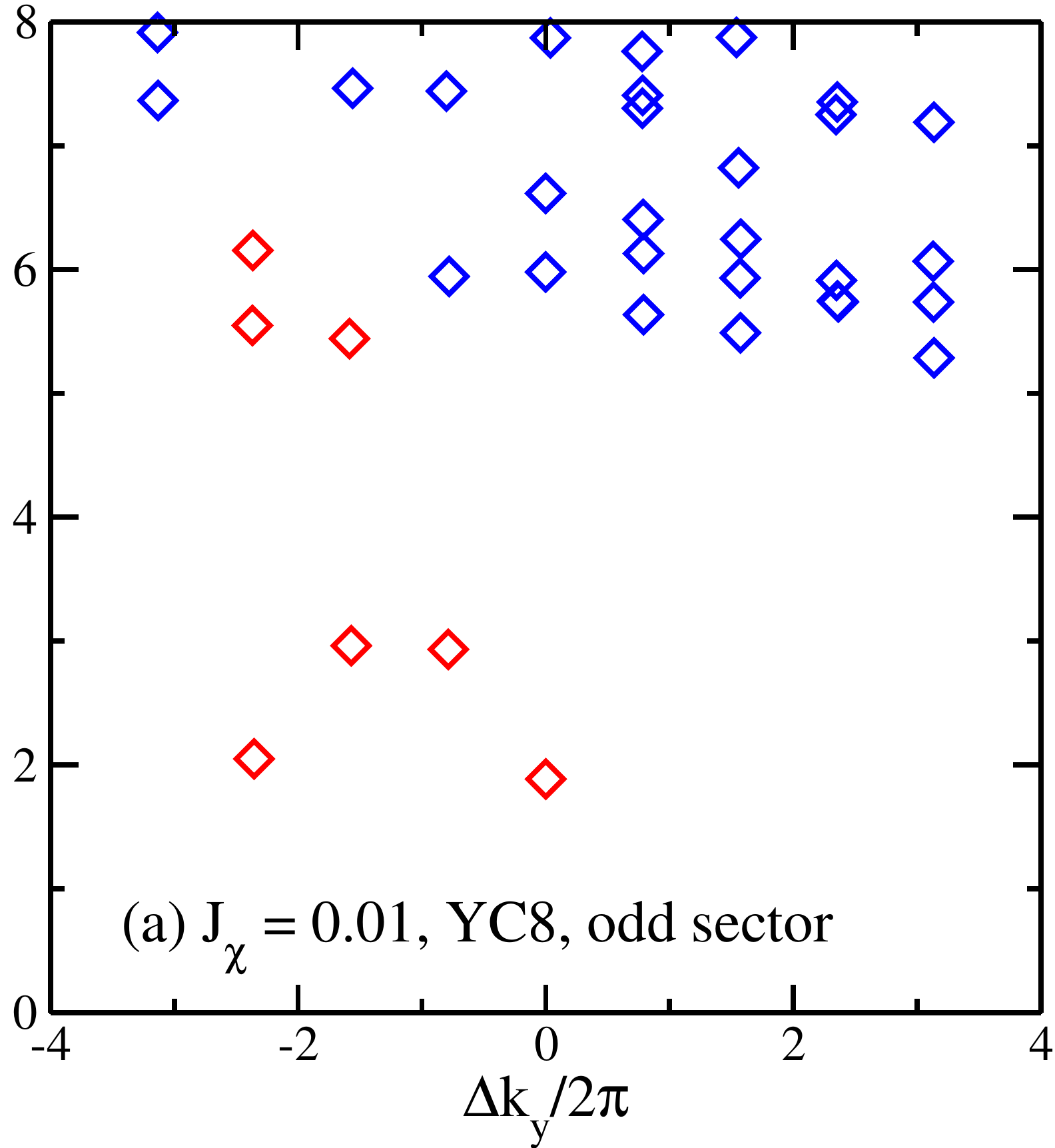}
\includegraphics[width=.45\linewidth]{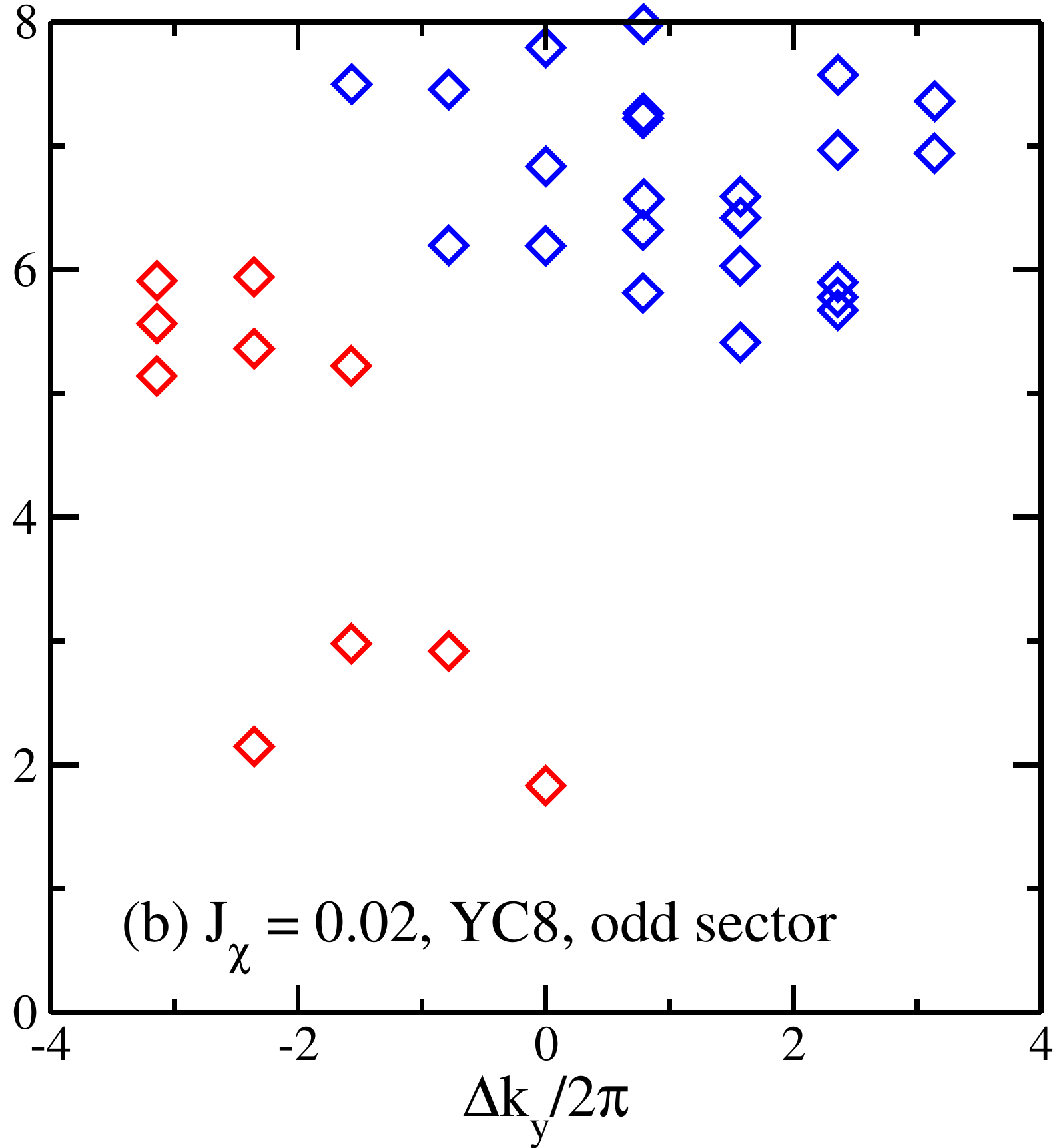}
\includegraphics[width=.46\linewidth]{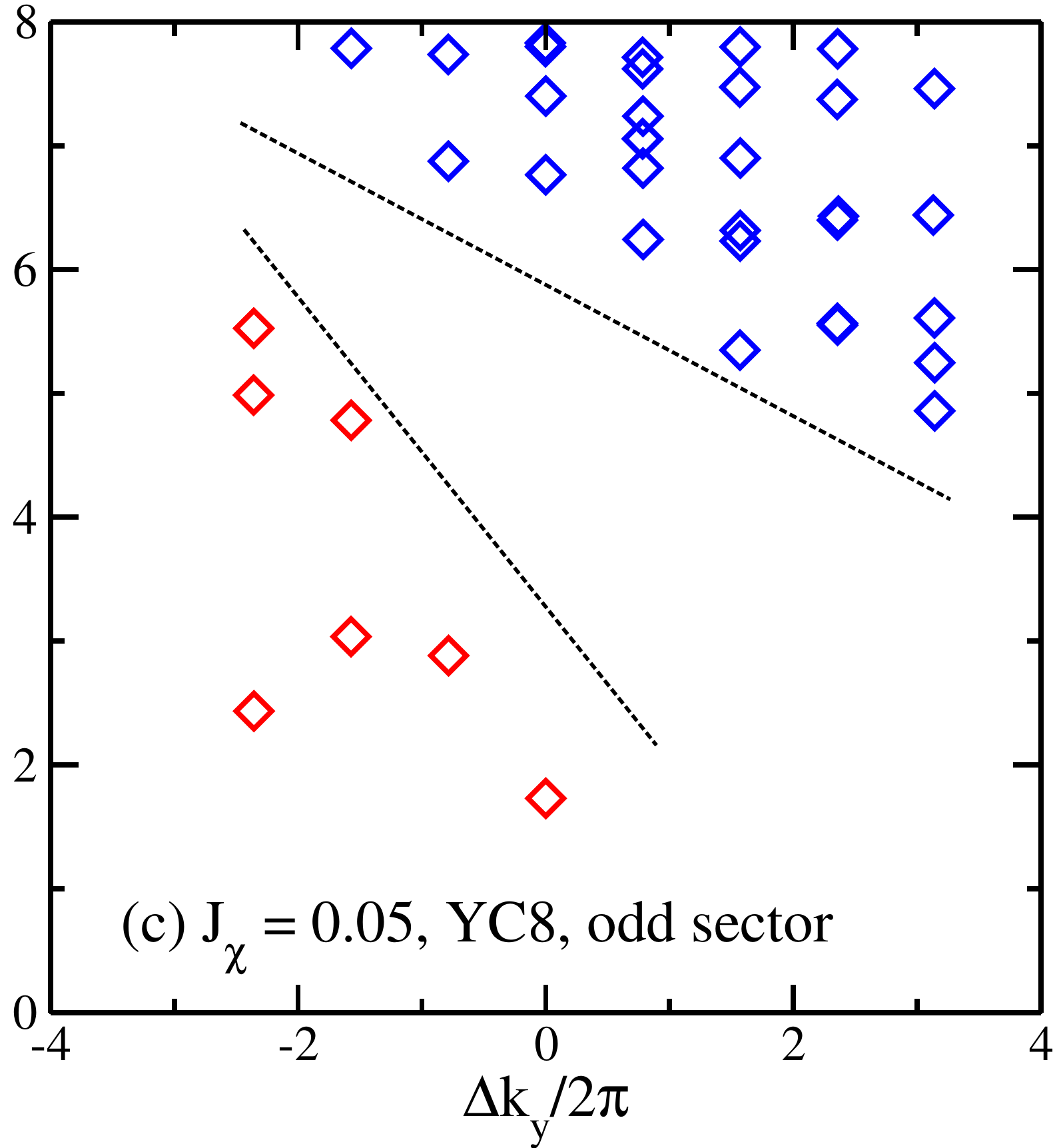}
\includegraphics[width=.46\linewidth]{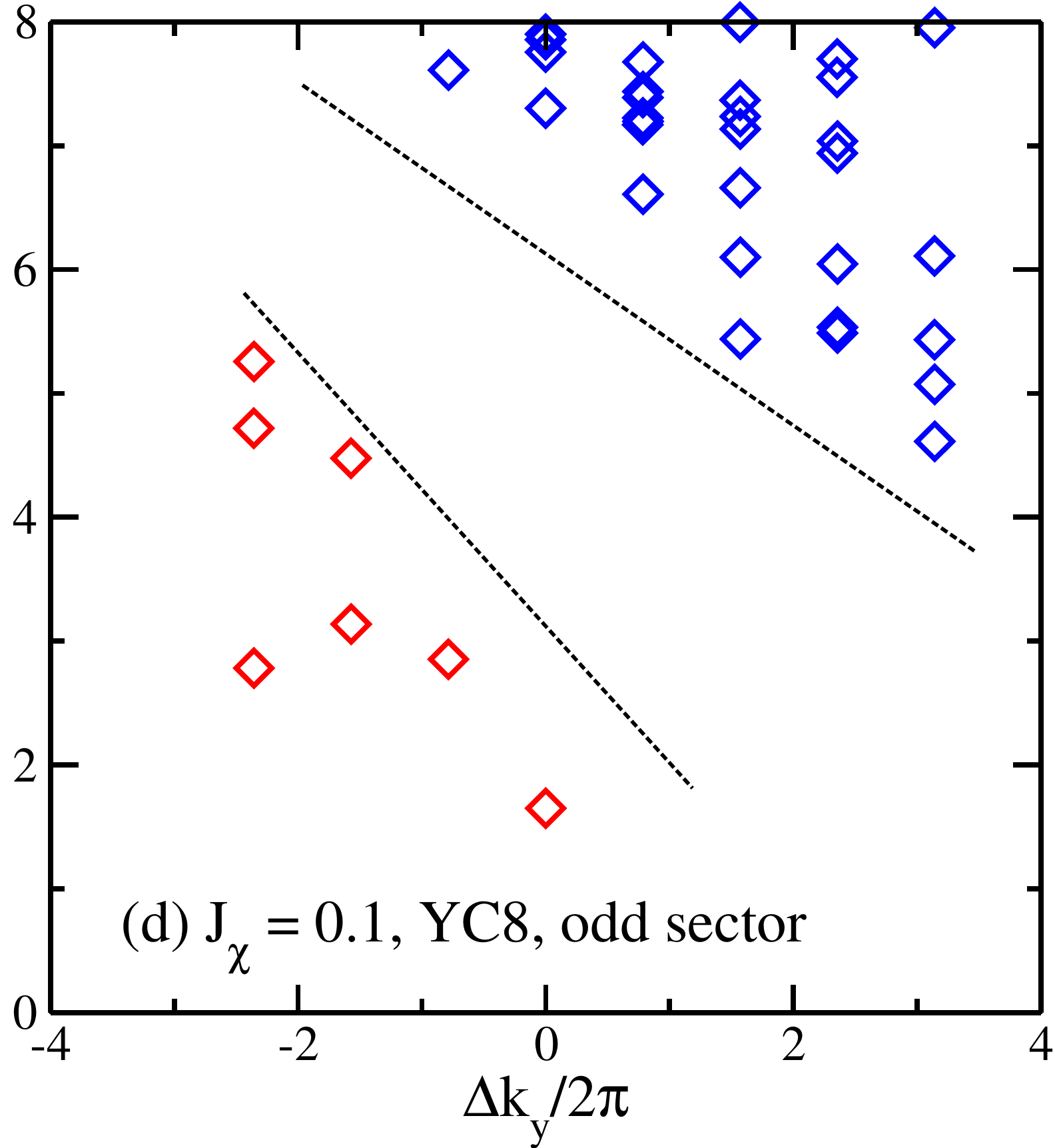}
\caption{Entanglement spectra of the ground state in the odd sector for $J_2 = 0.1$ and
different $J_{\chi}$ on the YC8-24 cylinder. $\Delta k_y$ is the relative momentum quantum number along 
the $y$ direction. $y$ label denotes the eigenvalue of the reduced density matrix $-\ln \lambda_i$. 
Here we show the spectra for total spin $S^z = 0$ sector.
}\label{spectrum}
\end{figure}

\begin{figure}[t]
\includegraphics[width=.45\linewidth]{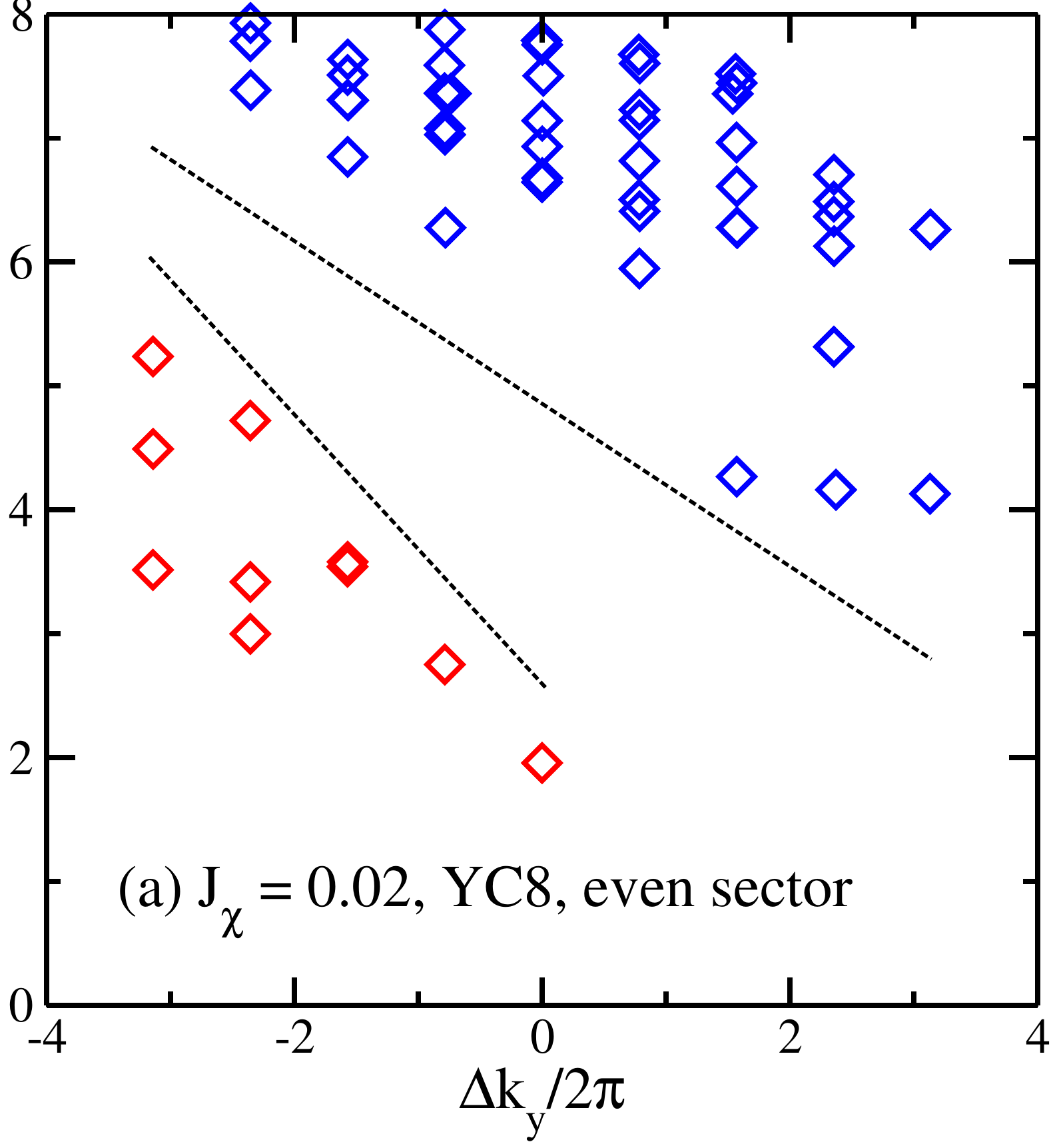}
\includegraphics[width=.45\linewidth]{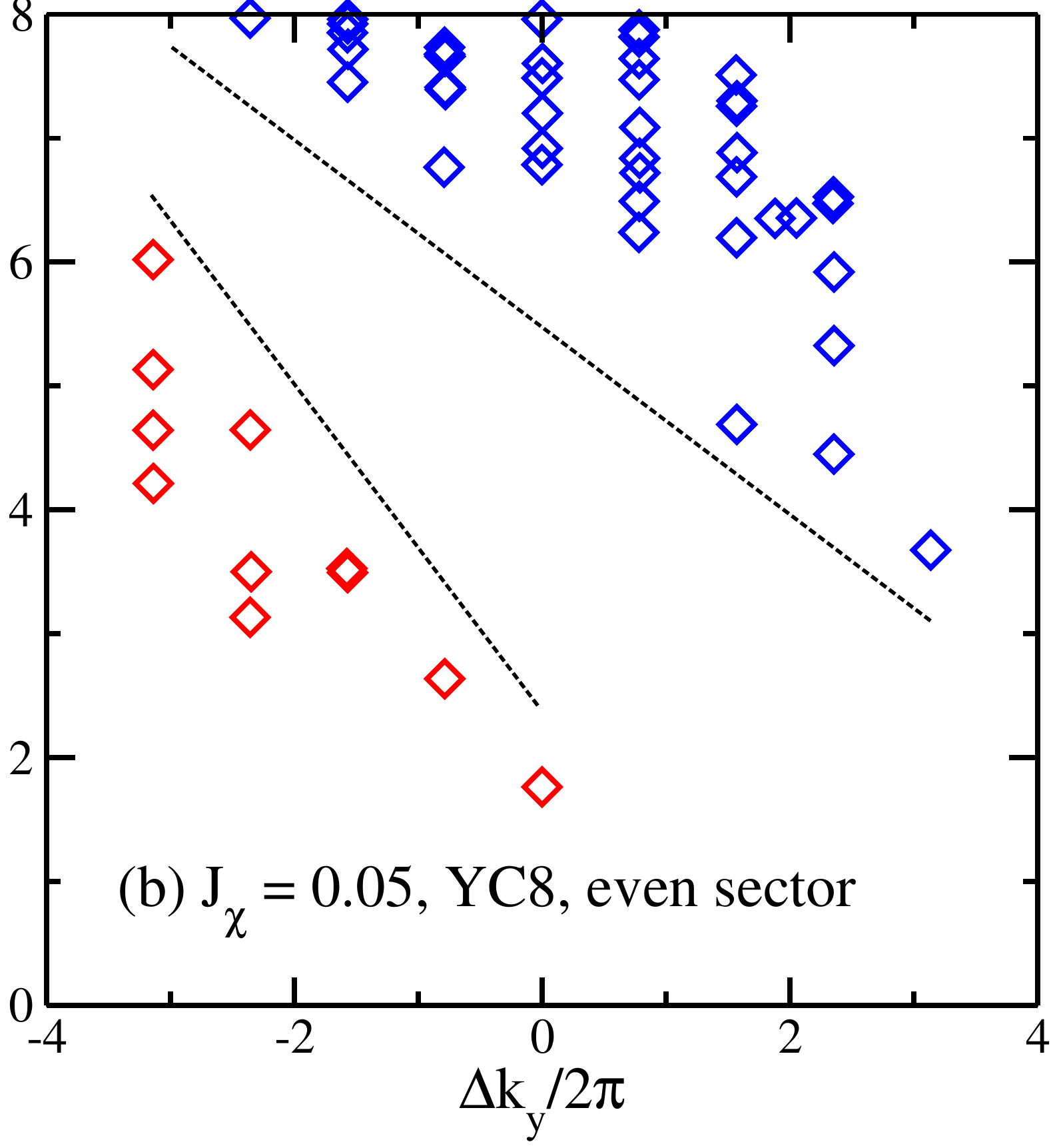}
\caption{Entanglement spectra of the ground state in the even sector for $J_2 = 0.1$ and
different $J_{\chi}$ on the YC8-24 cylinder. The labels are the same as Fig.~\ref{spectrum}.
}\label{spectrum_even}
\end{figure}

\begin{figure}[t]
\includegraphics[width=.9\linewidth]{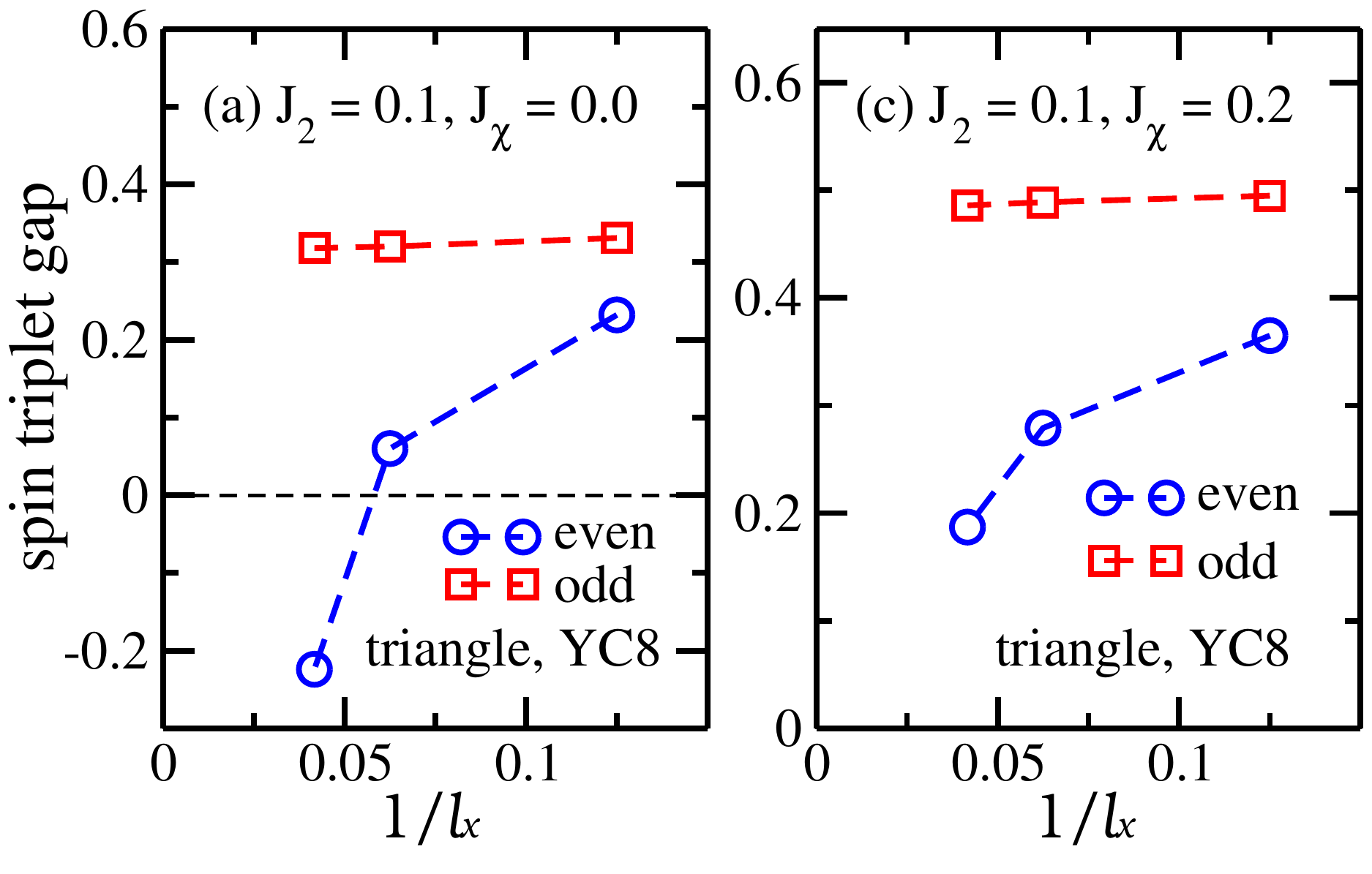}
\includegraphics[width=.9\linewidth]{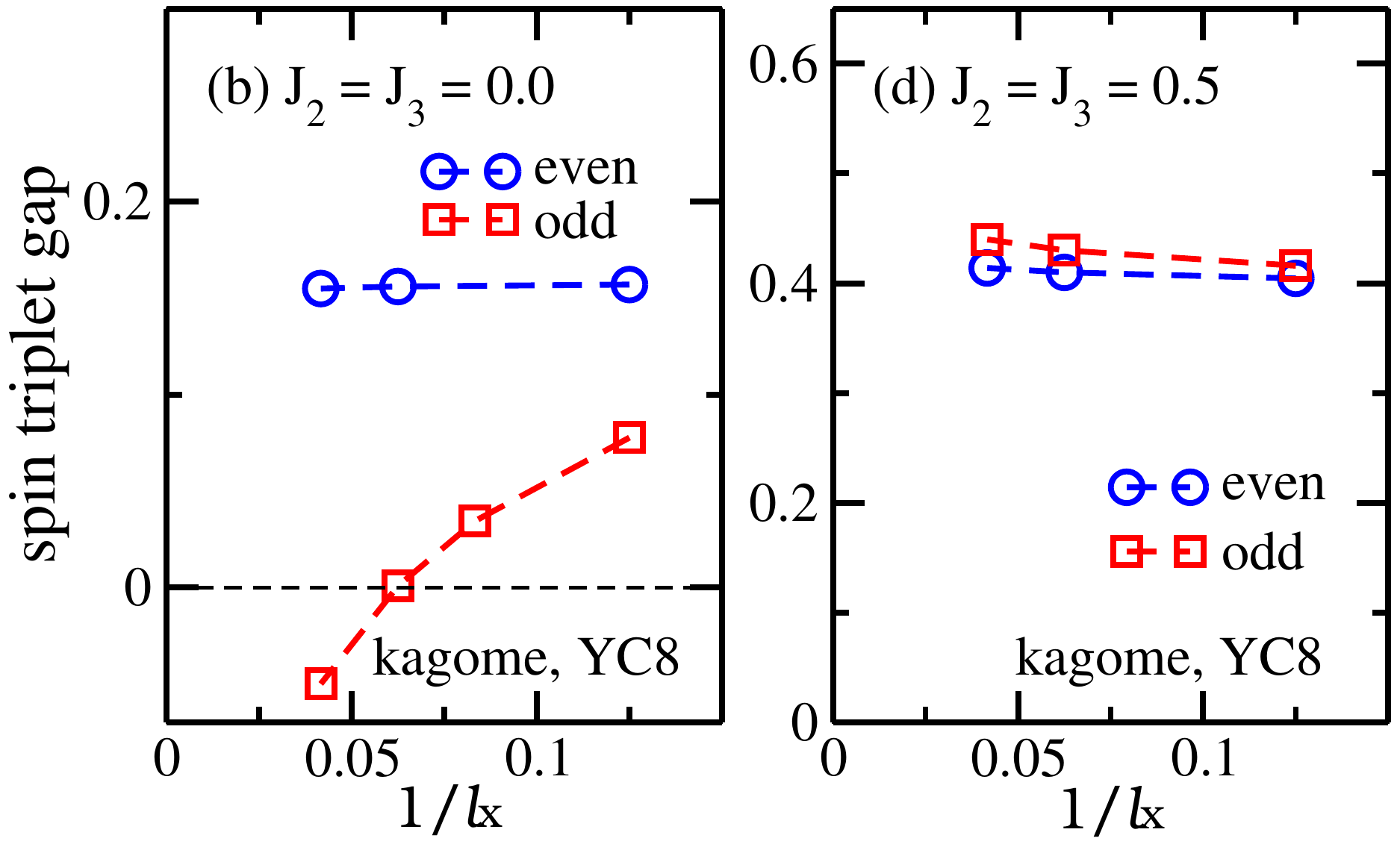}
\caption{Spin triplet gap versus cylinder length $l_x$ in the triangular and kagome models. 
We obtain the spin triplet gap by calculating the ground state on long cylinder first and then sweeping
the bulk sites for the ground state in spin quantum number $S = 1$ sector for a given length $l_x$. 
(a) and (c) are the $1/l_x$ dependence of the gap on the YC8 cylinder in the $J_1 - J_2$ 
SL ($J_2 = 0.1, J_{\chi} = 0.0$) and CSL ($J_2 = 0.1, J_{\chi} = 0.2$) for the triangular model. 
(b) and (d) are the $1/l_x$ dependence of the gap on the YC8 cylinder in the kagome spin liquid
($J_2 = J_3 = 0.0$) and CSL ($J_2 = J_3 = 0.5$) for the kagome model.
}\label{gap}
\end{figure}

Now we study the quantum phase transition from the $J_1 - J_2$ SL
to the CSL. We choose $J_2 = 0.1$ and switch on the chiral interaction $J_{\chi}$. 
In Fig.~\ref{transition}(a), we show the ground-state energy on the YC8 cylinder 
as well as on the $6 \times 6$ torus. The energy varies smoothly with growing $J_{\chi}$,
and we notice the slight change of energy for $J_{\chi} \lesssim 0.02$. Then we
compute the spin triplet gap $\Delta_{\rm T}$ on the YC8 cylinder based on the 
ground state with the lowest energy, which is in the odd sector as shown in 
Fig.~\ref{degeneracy}. The triplet gap is obtained by sweeping the total spin-1 
sector in the bulk of long cylinder~\cite{Yan2011}. We compare the obtained spin 
triplet gap by sweeping the spin-1 sector on the different system lengths, and we 
find the well converged triplet gap (one example can be found as the red square in Fig.~\ref{gap}(c)). 
In Fig.~\ref{transition}(b), we show the gap obtained by sweeping the middle $8\times 16$ 
sites in the spin-1 sector based on the ground state in the odd sector on the YC8-24 cylinder. 
The triplet gap changes slightly for $J_{\chi} \lesssim 0.02$. Above $J_{\chi} \simeq 0.02$, 
the gap grows fast, consistent with the non-zero gap in the CSL phase. The $J_{\chi}$ 
dependence of energy and triplet gap imply a possible phase transition at small $J_{\chi}$.

Next, we study the entanglement spectrum. As shown in Figs.~\ref{spectrum}(a)-(b) 
for $J_{\chi} = 0.01, 0.02$ in the odd sector with total spin $S^z = 0$, the ES
exhibit some features of the ES for $J_{\chi} = 0.0$~\cite{mcculloch2016}, where 
four eigenvalues are found below the higher spectrum. We also notice that with 
increasing $J_{\chi}$, some eigenvalues in the higher spectrum are decreasing
gradually as marked by red in Figs.~\ref{spectrum}(a)-(b). For $J_{\chi} = 0.05, 0.1$ 
as shown in Figs.~\ref{spectrum}(c)-(d), the decreasing eigenvalues seem to merge with the low-lying
levels, which are separated by an ES gap from the higher spectrum. The ES levels below the
gap exhibit the near degenerate pattern $\{1, 1, 2, 3\}$, which is consistent with 
the Laughlin CSL. The entanglement spectrum also suggests a phase transition at small 
$J_{\chi}$, which agrees with the transition suggested by energy and triplet gap in 
Fig.~\ref{transition}. In the even sector for $J_{\chi} = 0.0$, 
the low-lying part of the ES shows a deformed two-spinon continuum structure~\cite{mcculloch2016}. 
By switching on the chiral interaction, the low-lying part of the ES quickly changes 
to the structure that looks like the one in the CSL, which are shown in Fig.~\ref{spectrum_even}
and may suggest a stronger tendency to the chiral state in the even sector.

We also compute the spin triplet gap. For a comparison, 
we demonstrate the results of the same calculation for the spin-$1/2$ $J_1 - J_2 - J_3$ kagome model. 
Here, we obtain the triplet gap by calculating the ground state on 
long cylinder first (for example YC8-40) and then sweeping the bulk sites in the total 
spin $S = 1$ sector for the given bulk length $l_x$~\cite{Yan2011}. For
the $J_1 - J_2$ triangular model shown in Fig.~\ref{gap}(a), the spin gap measured
from the overall ground state (in the odd sector) is robust with increasing $l_x$.
However, the gap measured from the ground state in the even sector decreases
with $l_x$. On large $l_x$, the ground state in the even sector has the higher energy 
than the obtained state in the spin-1 sector, which suggests that the triplet gap
is lower than the total energy difference between the two sectors and is unlikely to support 
a well established gapped spin liquid. For the $J_1$ kagome model ($J_2 = J_3 = 0.0$) as 
shown in Fig.~\ref{gap}(b) on the YC8 cylinder, while the triplet gap measured from the 
overall ground state (in the even sector for the kagome spin liquid) is quite robust (consistent 
with the previous result~\cite{Yan2011, Depenbrock2012}), the gap in the odd sector decreases 
fast with $l_x$ and tends to vanish, seemly similar to the behaviors found in Fig.~\ref{gap}(a).

In the CSL phase as shown in Fig.~\ref{gap}(c), the overall triplet gap is also robust; 
but the gap in the even sector still decreases with $l_x$. In the well established CSL phase, 
for example the CSL phase in the kagome model as shown in Fig.~\ref{gap}(d)~\cite{gong2014kagome}, 
one can find the robust triplet gap in both sectors. The decreasing gap of the triangular 
CSL in the even sector suggests that on our studied system size the topological nature in 
the even sector is not fully developed. A possible reason is that this CSL regime is 
very close to the phase boundaries from the CSL to the neighboring phases.

\section{Quantum phase diagram}

\begin{figure}[t]
\includegraphics[width=0.7\linewidth]{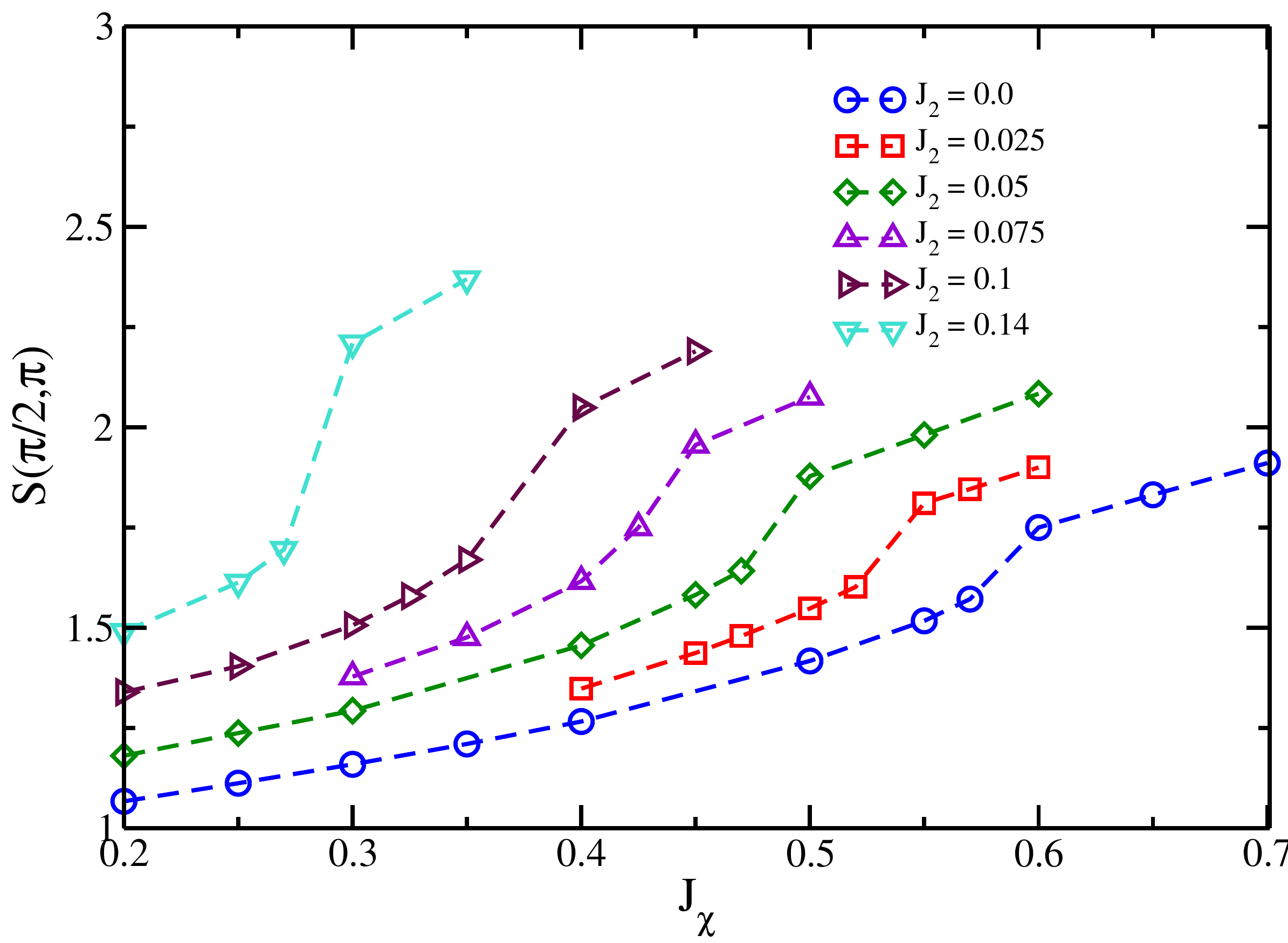}
\caption{$J_{\chi}$ dependence of the tetrahedral magnetic structure factor $S(\pi/2, \pi)$ 
on the YC8 cylinder for different $J_2$ couplings. The structure factor is obtained 
from the Fourier transform of the spin correlations of the middle $8\times 12$ sites on the 
YC8-24 cylinder. With increasing $J_{\chi}$, tetrahedral order shows a sharp increase that 
characterizes the phase transition to the tetrahedral phase.
}
\label{m_tetra}
\end{figure}

First of all, we study the phase transition from the tetrahedral phase to the CSL phase
by calculating magnetic structure factor. In Fig.~\ref{m_tetra}, we show the $J_{\chi}$ 
dependence of the tetrahedral structure factor peak on the YC8 cylinder, which is at 
$\vec{q} = (\pi/2, \pi)$. With increasing $J_{\chi}$, $S(\pi/2, \pi)$ shows a jump that 
characterizes the phase transition. Above the transition $J_{\chi}$, we find that the spin 
correlations decay quite slowly, consistent with a magnetic order developed. On the smaller 
YC6 and XC8 cylinders, the tetrahedral structure factor peak appears to increase smoothly,
which may be owing to the finite-size effects. We show the phase boundary in Fig.~\ref{fig:phase} 
based on the results on the larger YC8 cylinder.

\begin{figure}[t]
\includegraphics[width=0.45\linewidth]{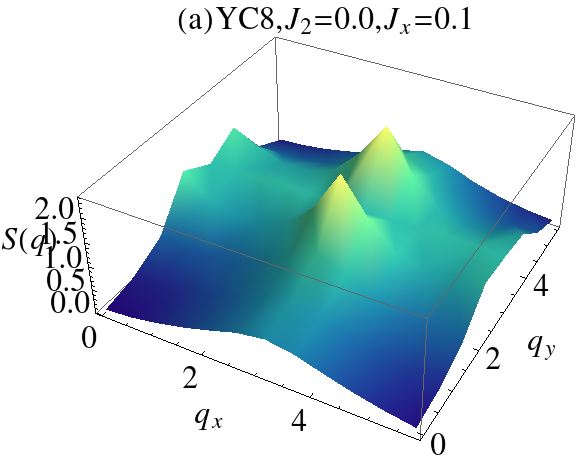}
\includegraphics[width=0.45\linewidth]{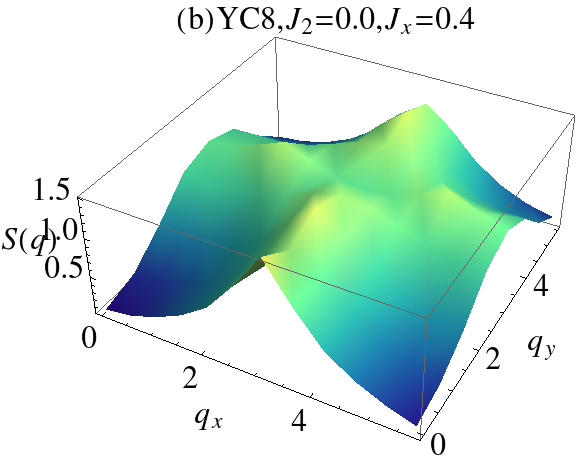}
\includegraphics[width=0.45\linewidth]{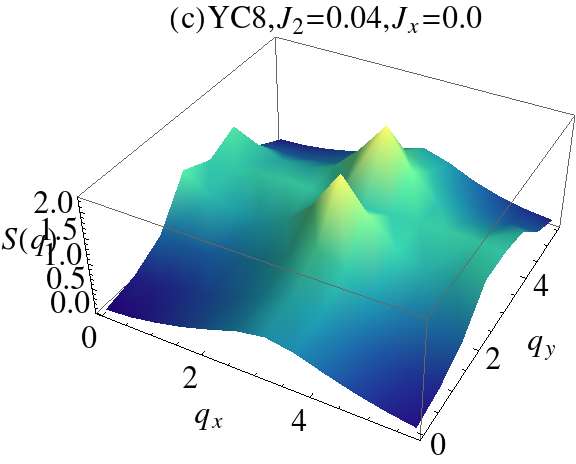}
\includegraphics[width=0.45\linewidth]{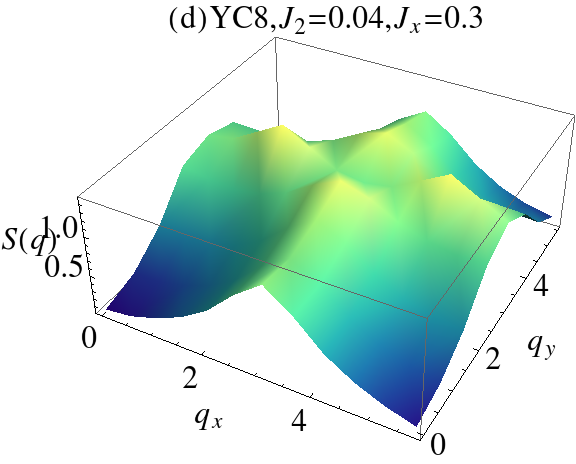}
\caption{Spin structure factor $S(\vec{q})$ in the $120^{\circ}$ magnetic order phase and the non-magnetic 
CSL phase. The structure factor is obtained from the middle $8\times 12$ sites on the YC8-24 cylinder. In 
the ordered phase for (a) and (c), $S(\vec{q})$ has the peak at $\vec{q} = (\pi, 2\pi/3)$. In the CSL phase 
for (b) and (d), $S(\vec{q})$ is featureless.
}
\label{tran_sq}
\end{figure}

By increasing the chiral interaction $J_{\chi}$ in the $120^{\circ}$ phase for $J_2 \lesssim 0.08$, 
the magnetic order is suppressed, leading to a transition from the magnetic order to the non-magnetic 
spin liquid phase. On our studied system size, we do not find sharp features to characterize this transition.
Thus we estimate a qualitative phase boundary as shown in Fig.~\ref{fig:phase}(c) by comparing
spin correlation function and magnitude of spin structure factor on the YC8 cylinder with the results at 
$J_2 = 0.08, J_{\chi} = 0$, where the system has the transition from the $120^{\circ}$ phase to the 
$J_1 - J_2$ SL phase. We roughly take the parameter points which have the similar magnitudes
of spin correlations and spin structure factor compared with those at $J_2 = 0.08, J_{\chi} = 0$ as 
the phase boundary. In Fig.~\ref{tran_sq}, we show the spin structure factor $S(\vec{q})$, which 
characterizes the present and absent magnetic order in the two phases.

\begin{figure}[t]
\includegraphics[width=0.7\linewidth]{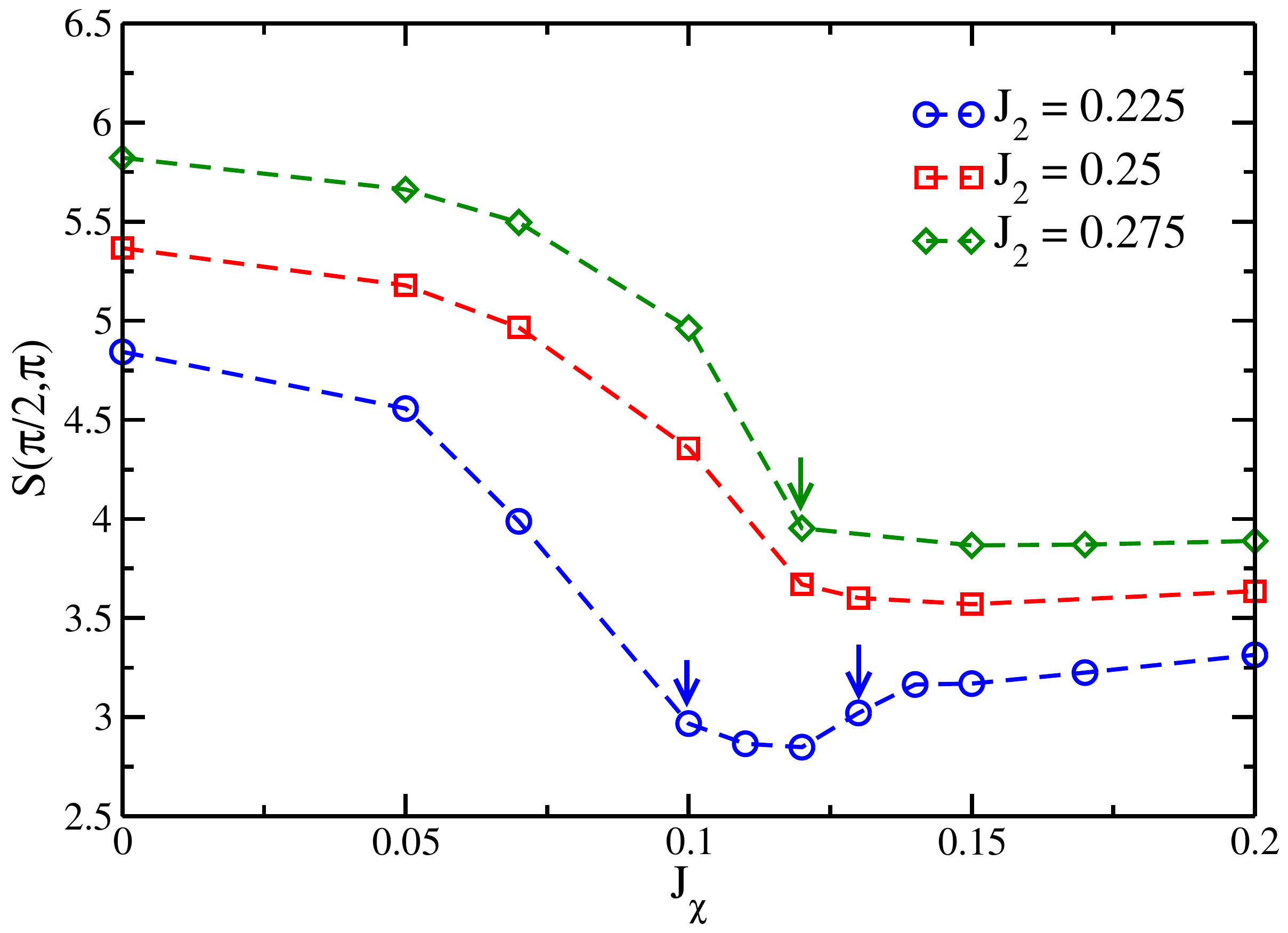}
\caption{$J_{\chi}$ coupling dependence of the magnetic structure factor at the momentum $\vec{q} = (\pi/2, \pi)$
for $J_2 = 0.225, 0.25, 0.275$ on the YC8-24 cylinder. The structure factor is obtained from the spin correlations
of the middle $8\times 12$ sites. The arrows denote the phase transitions. For $J_2 = 0.225$, we find two 
transitions with a CSL phase in between. For $J_2 = 0.25$ and $0.275$, we find a direct phase transition 
from the stripe phase to the tetrahedral phase.
}
\label{stripe_tetra}
\end{figure}

As shown in Fig.~\ref{fig:phase}(c), the stripe phase goes to the tetrahedral phase with 
or without an intermediate CSL phase depending on $J_2$. We calculate 
the spin correlations and structure factor on the YC8 cylinder. In both the stripe and the 
tetrahedral phase, spin structure factor shows the peak at $\vec{q} = (\pi/2, \pi)$.
For the case with the intermediate CSL phase, we find that $S(\pi/2, \pi)$ first decreases 
with growing $J_{\chi}$ and then keeps small in the CSL phase; with further increasing $J_{\chi}$, 
$S(\pi/2, \pi)$ shows a jump at the transition to the tetraheral phase, which is shown for 
$J_2 = 0.225$ in Fig.~\ref{stripe_tetra}. On the other hand, for $J_2 \gtrsim 0.25$,
$S(\pi/2, \pi)$ decreases fast with $J_{\chi}$ in the stripe phase and changes slowly in the 
tetrahedral phase, showing a kink to characterize the phase transition. We remark that on the
smaller YC6 cylinder for the $J_1 - J_2$ model, the magnetic order is weak at
$0.15 \lesssim J_2 \lesssim 0.2$; however, on the larger YC and XC cylinders the stripe order 
is quite robust. We demonstrate our phase diagram based on the large-size results.

\section{Summary and discussion}

We have studied the competing quantum phases of the spin-$1/2$ triangular $J_1-J_2$ Heisenberg model
with the additional scalar chiral interaction $J_{\chi}$ for $0 \leq J_2 \leq 0.3$ and 
$0 \leq J_{\chi} \leq 1.0$ by DMRG simulations. As shown in Fig.~\ref{fig:phase}(c), 
we find five phases: a non-coplanar tetrahedral magnetic order phase, a $120^\circ$ order phase, 
a stripe order phase, a $J_1 - J_2$ SL phase, and a chiral spin liquid (CSL) phase. 
The CSL is identified as the $\nu = 1/2$ bosonic fractional quantum Hall state, 
which seems to arise as a result of quantum fluctuations around the phase 
boundaries of the classical magnetic orders. In particular, the CSL state exhibits a strong bond 
anisotropy, strongly suggesting a nematic CSL with coexisting topological order and nematic order. 
We argue that the nematicity can be understood as a partially melted stripe order.
The emergent CSL induced from quantum fluctuations around classical phase boundaries has also been 
found in the kagome~\cite{gong2015, zhu2016, lauchli2015} and honeycomb~\cite{hickey2016} models, 
which may represent a common mechanism to generate novel quantum phases from frustration. We remark that 
while the analytic analyses suggest that the CSL is unlikely to emerge in the triangular model 
built out of weakly coupled chains with chiral interaction~\cite{sela2015}, DMRG results 
indicate that the CSL can emerge in the strong coupling regime.

By measuring the spin triplet gap and entanglement spectrum on the YC8 cylinder, we find a transition 
from the $J_1 - J_2$ SL to the CSL at small chiral coupling.
We also compute the spin triplet gap on cylinder geometry. While the gap above the overall ground
state (in the odd sector) is robust, the one in the even sector seems to be small in both spin liquid
phases. In the CSL phase, the small gap suggests that the even sector may be not well developed on
our studied system sizes. To find a robust CSL in finite-size calculation, other interactions may be
needed besides the chiral interaction. In the $J_1 - J_2$ SL, the vanishing gap in the even sector also
indicates that for a gapped spin liquid scenario the even sector is not well established. On the other
hand, in the gapless spin liquid scenario~\cite{iqbal2016}, one finds it inconsistent with the 
odd sector exhibiting a large gap based on DMRG calculations ($\Delta_{\rm T} \simeq 0.35J_1$ on the 
YC8 cylinder and the size scaling seems to suggest a large gap in the large-size
limit~\cite{zhuzhenyue2015,Hu2015}). However, the spin gap measured in DMRG
calculations may also have large finite-size effects, which requires further investigations.
We hope that future sutdies will be able to resolve between these scenarios more clearly.
\\

This research is supported by the state of Florida (S.S.G.),
National Science Foundation Grants DMR-1157490 (S.S.G. and K.Y.), DMR-1442366 (K.Y.), and DMR-1408560 (D.N.S.).
Work at Los Alamos was supported by U.S. DOE National Nuclear Security Administration through Los Alamos
National Laboratory LDRD Program (W.Z. and J.-X.Z.), and in part supported by the Center for Integrated
Nanotechnologies, a U.S. DOE Basic Energy Sciences user facility.
\\

{\it Note Added.---} While completing this work, we became aware of
a related paper~\cite{saadatmand2017} that also studies the robustness
of the $J_1 - J_2$ spin liquid against the chiral interaction in the
same $J_1 - J_2 - J_{\chi}$ triangular model. We find the overall agreement
with Ref.~\onlinecite{saadatmand2017}.

\bibliography{csl_tri}

\end{document}